\documentclass[10pt]{article}
\usepackage{amsfonts}
\usepackage{indentfirst}
\usepackage{booktabs}
\usepackage{cite}
\usepackage{amssymb,amsmath,mathrsfs}
\usepackage{caption,subfigure,graphicx}
\usepackage{subfigure}
\usepackage{extarrows}
\begin{document}
\title{\bf{A coupled focusing-defocusing complex short pulse equation: multisoliton, breather, and rogue wave } }
\author{ Jun Yang, Zuo-Nong Zhu
\footnote{Corresponding author. Email: znzhu@sjtu.edu.cn}\\
School of Mathematical Sciences, Shanghai Jiao Tong University,\\ 800 Dongchuan Road, Shanghai, 200240, P. R. China\\
}
\date{ }
\maketitle
\begin{abstract}
 Nonlinear Schr\"odinger equation, short pulse
equation and complex short pulse
equation have important application in nonlinear optics. They can be derived from the Maxwell equation. In this paper, we investigate a coupled focusing-defocusing complex short pulse equation.
 The bright-bright, bright-dark and dark-dark soliton solutions of the coupled focusing-defocusing complex short pulse equation are given.
 Then the breathers are derived from the dark soliton solution. The rogue wave solutions are also constructed.
 The dynamics and the asymptotic behavior of the soliton solutions are analyzed,
 which reveals that there exist the elastic or inelastic collision
in bright-bright soliton solution. But the interactions of bright-dark and dark-dark soliton solutions
are both elastic.
\end{abstract}
\section{Introduction}
As is well known,  nonlinear Schr\"odinger(NLS) equation and the coupled NLS equation play an
important role in nonlinear optics, plasma physics, water waves, and Bose-Einstein condensates \cite{1,2,3,4,5,6,7}.
However, in optical fibers, the NLS equation becomes less accurate to
describe the propagation of ultra-short pulses because the width
of optical pulse is in the order of femtosecond$(10^{-15}s)$\cite{8}.
 Sch\"afer and Wayne derived the short pulse
(SP) equation\cite{9}
\begin{equation}\label{1}
u_{xt}=u +\frac{1}{6}(u^{3})_{xx},
\end{equation}
which can describe the propagation of ultra-short optical pulses in nonlinear
media. The short pulse equation is completely integrable\cite{10,11}. Connection
between the short pulse equation and the sine-Gordon equation has been given by a hodograph transformation\cite{12,13}.
Multi-soliton, multi-breather and rogue wave solutions for Eq.\eqref{1} have been constructed\cite{14,15,16}.
The geometric interpretation and integrable discretization of Eq.\eqref{1} have also been studied\cite{17,18,19}.
It has been revealed that there are great advantages
of dealing with the complex short pulse(CSP)
equation and complex multi-component short pulse
equations which can more appropriately describe
the propagation of optical pulse along the optical
fibers\cite{20,21,22}.
The complex short pulse equation\cite{20,22} is
\begin{equation}\label{2}
q_{xt}+q +\frac{1}{2}\sigma(|q|^{2}q_{x})_{x}=0,~~
\end{equation}
where $\sigma=\pm1$ represents focusing- and defocusing-type. Its two-component form
\begin{align}\label{3}
q_{1,xt}+q_{1} +\frac{1}{2}((|q_{1}|^{2}+|q_{2}|^{2})q_{1,x})_{x}=0, \\  \notag
q_{2,xt}+q_{2} +\frac{1}{2}((|q_{1}|^{2}+|q_{2}|^{2})q_{2,x})_{x}=0,
\end{align}
is first proposed by Feng\cite{20}. It can be derived from the Maxwell equation\cite{20}.
The multi-bright-soliton, multi-breather and higher-order
rogue wave solution of Eq.\eqref{2} for focusing case is derived in\cite{20}.
Meanwhile, the multi-dark-soliton of Eq.\eqref{2} for defocusing case is studied in\cite{22}.
For the two-component focusing-focusing case\eqref{3}, the bright-bright and bright-dark soliton
solutions are obtained \cite{20,23}.
To the best of our knowledge, the focusing-defocusing case for the coupled
complex short pulse equation has not been studied.
In this paper, we investigate the following
coupled focusing-defocusing complex short pulse equation:
\begin{align}\label{4}
&q_{1,xt}+q_{1} +\frac{1}{2}((\sigma_{1}|q_{1}|^{2}+\sigma_{2}|q_{2}|^{2})q_{1,x})_{x}=0, \\ \notag
&q_{2,xt}+q_{2} +\frac{1}{2}((\sigma_{1}|q_{1}|^{2}+\sigma_{2}|q_{2}|^{2})q_{2,x})_{x}=0,
\end{align}
where $\sigma_{1}=-\sigma_{2}=1$. The coupled focusing-defocusing complex short pulse equation was introduced by Feng \cite{24}. Starting from the Lax pair of the coupled focusing-defocusing
complex short pulse equation, we will construct its soliton solutions, including
the bright-bright soliton, bright-dark soliton, dark-dark soliton,
breather and rogue wave solutions.
The asymptotic behaviors of two soliton solutions are studied. It will be shown that the coupled
bright-bright soliton solution undergo fascinating interactions with redistribution of energy.
We construct bright-dark soliton solution and dark-dark soliton solution.
This means the bright-bright soliton and dark-dark soliton are supported simultaneously
for the coupled focusing-defocusing complex short pulse equation.
The collision of the two solitons are completely elastic behaviors.
Furthermore, the breather solution is constructed and the rogue wave soliton solution
is derived by resorting to the Taylor series expansion coefficients of the breather solutions.
\section{Lax pairs and N-bright soliton solutions for the coupled focusing-defocusing complex short pulse equation}
\textbf{2.1 Lax pairs and bilinear form of Eq.\eqref{4}}\\
The Lax pair for the coupled focusing-defocusing complex short pulse equation~\eqref{4} is
\begin{equation}\label{5}
\Psi_{x}=U\Psi,~~~~~~~~~\Psi_{t}=V\Psi,
\end{equation}
with
\begin{equation*}   \label{6}
 U=\lambda\left(
    \begin{array}{cc}
       I_{2} & Q_{x} \\
      R_{x} & -I_{2} \\
    \end{array}
  \right),
  ~~~~~~~~~~~~~~~~~~~
  V=\left(
    \begin{array}{cc}
       -\frac{\lambda}{2}QR-\frac{1}{4\lambda}I_{2} & -\frac{\lambda}{2}QRQ_{x}+\frac{1}{2}Q \\
       -\frac{\lambda}{2}RQR_{x}-\frac{1}{2}R &  \frac{\lambda}{2}QR+\frac{1}{4\lambda}I_{2} \\
    \end{array}
  \right),
\end{equation*}
where $I_{2}$ is a $2\times2$ identity matrix, $Q,~R$ are $2\times2$ matrices defined
as
\begin{equation}\label{7}
Q=\left(
     \begin{array}{cc}
    q_{1} & q_{2}\\    \notag
q_{2}^{*}&q_{1}^{*}\\  \notag
           \end{array}
   \right),~~~~~~~~~~~~~~~~~
   R=\left(
     \begin{array}{cc}
    q_{1}^{*} & -q_{2}\\    \notag
   -q_{2}^{*}&q_{1}\\  \notag
           \end{array}
   \right).
\end{equation}
Notice that
\begin{equation}\label{8}
 QR=RQ=(|q_{1}|^{2}-|q_{2}|^{2})I_{2},
\end{equation}
one can check that the compatibility condition $U_{t}-V_{x}+
[U,V]=0$ for \eqref{5} gives the coupled focusing-defocusing complex short pulse equation
\begin{align}\label{9}
q_{1,xt}+q_{1} +\frac{1}{2}((|q_{1}|^{2}-|q_{2}|^{2})q_{1,x})_{x}=0, \\  \notag
q_{2,xt}+q_{2} +\frac{1}{2}((|q_{1}|^{2}-|q_{2}|^{2})q_{2,x})_{x}=0.
\end{align}

The coupled focusing-defocusing complex short pulse equation can be bilinearized as
\begin{align}  \notag
&D_{s}D_{y}g\cdot f=gf, \\ \label{10}
&D_{s}D_{y}h\cdot f=hf, \\ \notag
&D_{s}^{2}f\cdot f=\frac{1}{2}(|g|^{2}-|h|^{2}),
\end{align}
by dependent variable transformation
 \begin{equation}
 q_{1}=\frac{g}{f},~~q_{2}=\frac{h}{f},
 \end{equation}
and the hodograph transformation
\begin{equation}\label{tran1}
x=y-2(\ln f)_{s},~~t=-s,
\end{equation}
where
\begin{equation}
D_{s}^{m}D_{y}^{n}f.g=(\partial_{s}-\partial_{s^{'}})^{m}(\partial_{y}-\partial_{y^{'}})^{n}f(y,s).g(y^{'},s^{'})|_{s^{'}=s,y^{'}=y}. \notag
\end{equation}
\textbf{Proof}: Dividing both sides by $f^{2}$ converts the bilinear equation\eqref{10} to
\begin{align}\label{11a}
(\frac{g}{f})_{sy}+2\frac{g}{f}(lnf)_{sy}=\frac{g}{f},\\\label{11b}
(\frac{h}{f})_{sy}+2\frac{h}{f}(lnf)_{sy}=\frac{h}{f},\\ \label{11c}
(lnf)_{ss}=\frac{1}{4}\frac{(|g|^{2}-|h|^{2})}{f^{2}}.
\end{align}
From the dependent variable transformation and hodograph transformation, we have
\begin{align}\label{12}
\frac{\partial x}{\partial s}=-2 (lnf)_{ss}=-\frac{1}{2}(|q_{1}|^{2}-|q_{2}|^{2}),~~~
\frac{\partial x}{\partial y}=1-2 (lnf)_{sy},
\end{align}
which implies
\begin{equation}\label{13}
\partial_{y}=\rho^{-1}\partial_{x}~~~~~~~
\partial_{s}=-\partial_{t}-\frac{1}{2}(|q_{1}|^{2}-|q_{2}|^{2})\partial_{x},
\end{equation}
where $1-2 (lnf)_{sy}=\rho^{-1}$.
The Eqs.(11) and (12) give
\begin{align} \label{16}
\partial_{x}(-\partial_{t}-\frac{1}{2}(|q_{1}|^{2}-|q_{2}|^{2})\partial_{x})q_{1}=q_{1},\\   \notag
\partial_{x}(-\partial_{t}-\frac{1}{2}(|q_{1}|^{2}-|q_{2}|^{2})\partial_{x})q_{2}=q_{2}.
\end{align}
\textbf{2.2 N-bright soliton solution}\\
N-soliton solution to the Eq.\eqref{9} can be expressed by the following pfaffians:
\begin{align}\label{17}
f&=(-1)^{N}{\rm Pf}(a_{1},\ldots,a_{2N},b_{1},\ldots, b_{2N}),\\  \notag
g&=(-1)^{N}{\rm Pf}(d_{0},\gamma_{1},a_{1},\ldots,a_{2N},b_{1},\ldots,b_{2N}),\\  \notag
h&=(-1)^{N}{\rm Pf}(d_{0},\gamma_{2},a_{1},\ldots,a_{2N},b_{1},\ldots,b_{2N}).  \notag
\end{align}
The elements of the pfaffians are determined as
\begin{align}\label{18}
{\rm Pf}(a_{j},a_{k})&=\frac{p_{j}-p_{k}}{p_{j}+p_{k}}e^{\eta_{j}+\eta_{k}},~~~{\rm Pf}(a_{j},b_{k})=\delta_{j,k} \\ \notag
{\rm Pf}(b_{j},b_{k})&=\frac{1}{4}\frac{\alpha_{j}\alpha_{k}-\beta_{j}\beta_{k}}{p_{j}^{-2}-p_{k}^{-2}}\delta_{\mu+1,\nu},~~~{\rm Pf}(d_{l},a_{k})=p_{k}^{l}e^{\eta_{k}}\\ \notag
{\rm Pf}(b_{j},\gamma_{1})&=\alpha_{j}\delta_{\mu,i},~~~~~{\rm Pf}(b_{j},\gamma_{2})=\beta_{j}\delta_{\mu,i},\\  \notag
{\rm Pf}(d_{0},b_{j})&={\rm Pf}(d_{0},\gamma_{1})={\rm Pf}(d_{0},\rho_{i})={\rm Pf}(a_{j},\gamma_{2})={\rm Pf}(a_{j},\gamma_{i})=0.
\end{align}
Here we need to define two sets:
$B_{\mu}(\mu=1,2): B_{1}=\{b_{1},b_{2}, \cdots,b_{N}\}, B_{2}=\{b_{N+1}, b_{N+2}, \cdots,b_{2N}\}$ and an
index function of $b_{j}$ by index$(b_{j})=\mu$ if $b_{j}\in B_{\mu}$. $\mu=$index$(b_{j})$, $\nu=$index$(b_{k})$, $\eta_{j}=p_{j}y+p_{j}^{-1}s+\eta_{j,0}$, which satisfy $p_{j+N}=\bar{p}_{j}$,~
$\alpha_{j+N}=\bar{\alpha}_{j}$,~$\beta_{j+N}=\bar{\beta}_{j}$, where $\bar{p}_{j}$,~$\bar{\alpha}_{j}$ and $\bar{\beta}_{j}$
represent the complex conjugates of $p_{j}$~$\alpha_{j}$ and $\beta_{j}$.\\

The proof of the N-soliton solution is similar to the ones of the coupled
focusing-focusing complex short pulse equation
with $\gamma_{j}$ substituting $\beta_{j} (j=1,2)$ in appendix (A.9)\cite{23} except the following formula
\begin{align*}
\frac{1}{2}(g_{1}g_{1}^{*}-g_{2}g_{2}^{*})=&\frac{1}{2}\sum_{i,j}^{2N}(-1)^{i+j}{\rm Pf}(b_{i},\gamma_{1})
{\rm Pf}(d_{0},\cdots,\hat{b}_{i},\cdots)\times {\rm Pf}(b_{j},\gamma_{1}^{*})
{\rm Pf}(d_{0},\cdots,\hat{b}_{j},\cdots)\\
&-\frac{1}{2}\sum_{i,j}^{2N}(-1)^{i+j}{\rm Pf}(b_{i},\gamma_{2})
{\rm Pf}(d_{0},\cdots,\hat{b}_{i},\cdots)\times {\rm Pf}(b_{j},\gamma_{2}^{*})
{\rm Pf}(d_{0},\cdots,\hat{b}_{j},\cdots)\\
=&\frac{1}{4}\sum_{i,j}^{2N}(-1)^{i+j}\alpha_{i}\alpha_{j}^{*}\times
{\rm Pf}(d_{0},\cdots,\hat{b}_{i},\cdots){\rm Pf}(d_{0},\cdots,\hat{b}_{j},\cdots)\\
-&\frac{1}{4}\sum_{i,j}^{2N}(-1)^{i+j}\beta_{i}\beta_{j}^{*}\times
{\rm Pf}(d_{0},\cdots,\hat{b}_{i},\cdots)
{\rm Pf}(d_{0},\cdots,\hat{b}_{j},\cdots)\\
=&\frac{1}{4}\sum_{i,j}^{2N}(-1)^{i+j}(\alpha_{i}\alpha_{j}^{*}
-\beta_{i}\beta_{j}^{*})\times {\rm Pf}(d_{0},\cdots,\hat{b}_{i},\cdots){\rm Pf}(d_{0},\cdots,\hat{b}_{j},\cdots)\\
=&\frac{1}{2}\sum_{i,j}^{2N}(-1)^{i+j}(p_{i}^{-2}-p_{j}^{-2})\times{\rm Pf}(b_{i},b_{j}){\rm Pf}(d_{0},\cdots,\hat{b}_{i},\cdots){\rm Pf}(d_{0},\cdots,\hat{b}_{j},\cdots).
\end{align*}
We can see that $q_{1}=\frac{g}{f},~q_{2}=\frac{h}{f}$ is the N-bright soliton
solutions to the Eq.\eqref{9} under the hodograph transformation \eqref{tran1},
where $f,g,h$ are described by Eq.\eqref{17}.

\textbf{ 2.2.2 One soliton solution}\\
\begin{figure}
\begin{minipage}{0.3\linewidth}\centering
\includegraphics[width=1\textwidth]{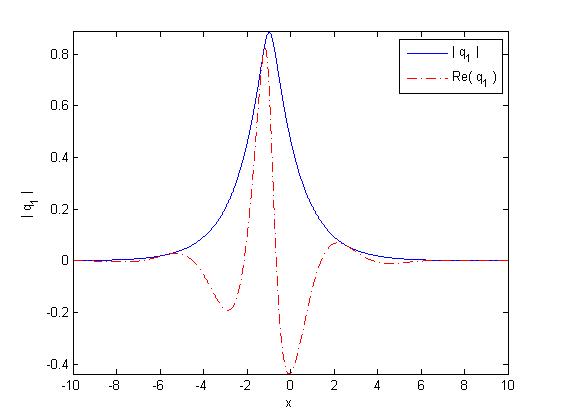}\caption*{(a)}
\end{minipage}
\begin{minipage}{0.3\linewidth}\centering
\includegraphics[width=1\textwidth]{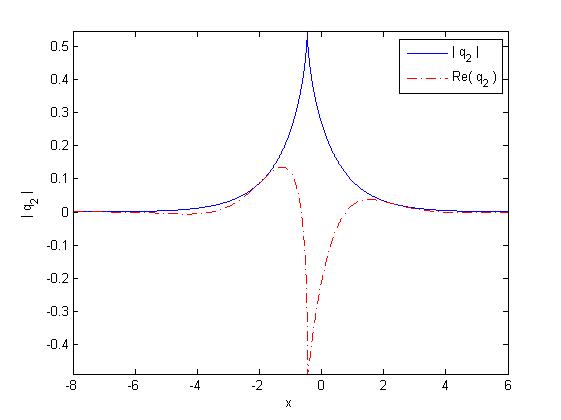}\caption*{(b)}
\end{minipage}
\begin{minipage}{0.3\linewidth}\centering
\includegraphics[width=1\textwidth]{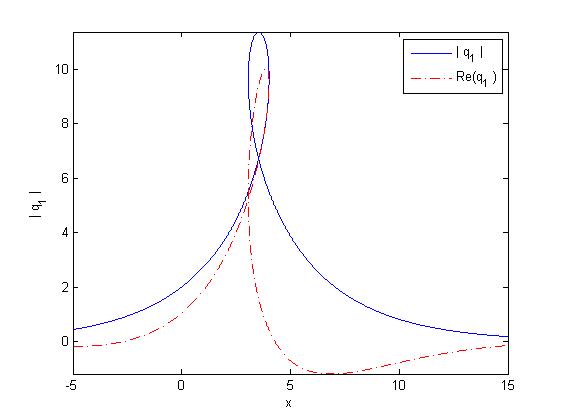}\caption*{(c)}
\end{minipage}
\caption{\small{One bright soliton to the coupled focusing-defocusing complex short pulse equation:
(a)-(c) profiles with solid line $|q|$ and dashed line $Re(q)$ at $t =0$.
(a) smooth soliton with $p_{1}=0.8+1.3i, \alpha_{1}=1+2i, \beta_{1}=1-i$,
(b) cuspon soliton with $p_{1}=1+i, \alpha_{1}=2-3i, \beta_{1}=1+1.4i$,
(c) loop soliton with $p_{1}=2-i, \alpha_{1}=1+1.7i, \beta_{1}=1-1.5i$.}}
\label{fig1}
\end{figure}
Based on \eqref{17}, one-soliton solution $q_{1}=\frac{g}{f},~q_{2}=\frac{h}{f}$ is given by
\begin{align}\label{19}
f&=1+\frac{1}{4}\frac{(|\alpha_{1}|^{2}-|\beta_{1}|^{2})|p_{1}|^{4}}{(p_{1}+p_{1}^{*})^{2}}e^{\eta_{1}+\eta_{1}^{*}}, \\
g&=\alpha_{1}e^{\eta_{1}},h=\beta_{1}e^{\eta_{1}}.
\end{align}

In order to obtain the bright soliton, the condition $|\alpha_{1}|>|\beta_{1}|$ needs to be satisfied.
Let $p_{1}=p_{\rm\scriptscriptstyle 1R}+ip_{\rm\scriptscriptstyle 1I}$ and assume $p_{\rm\scriptscriptstyle 1R}>0$
then the one-soliton solution is
\begin{align}\label{20}
\left(
  \begin{array}{c}
    q_{1} \\
    q_{2} \\
  \end{array}
\right)&=
\left(
  \begin{array}{c}
    A_{1} \\
    A_{2} \\
  \end{array}
\right)\frac{2p_{\rm\scriptscriptstyle 1R}}{|p_{1}|^{2}}e^{i\eta_{\rm\scriptscriptstyle 1I}}
\textup{sech}(\eta_{\rm\scriptscriptstyle 1R}+\eta_{10}),\\  \notag
x&=y-2\frac{p_{\rm\scriptscriptstyle 1R}}{|p_{1}|^{2}}
(1+\tanh(\eta_{\rm\scriptscriptstyle 1R}+\eta_{10})),~t=-s,
\end{align}
\begin{figure}[!h]
\centering
\begin{minipage}{0.3\linewidth}\centering
\includegraphics[width=1\textwidth]{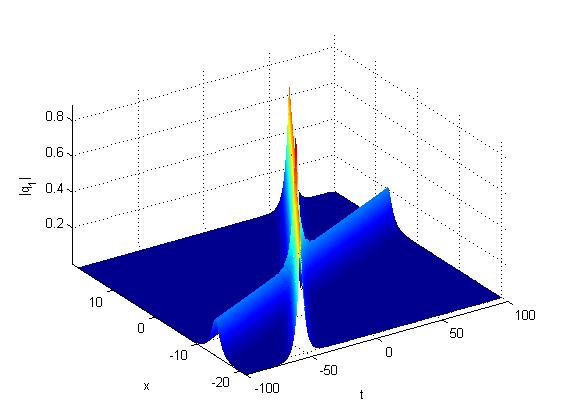}\caption*{(a1)}
\end{minipage}
\begin{minipage}{0.3\linewidth}\centering
\includegraphics[width=1\textwidth]{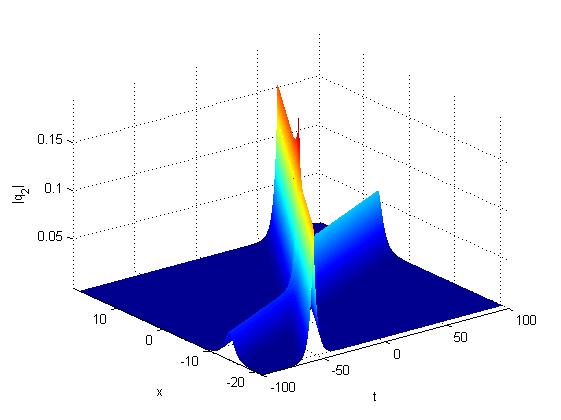}\caption*{(a2)}
\end{minipage}\\
\begin{minipage}{0.3\linewidth}\centering
\includegraphics[width=1\textwidth]{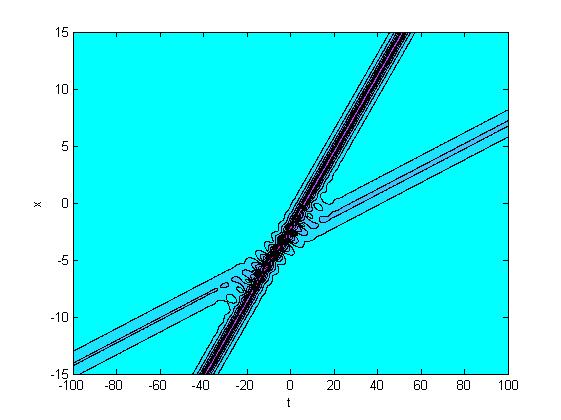}\caption*{(b1)}
\end{minipage}
\begin{minipage}{0.3\linewidth}\centering
\includegraphics[width=1\textwidth]{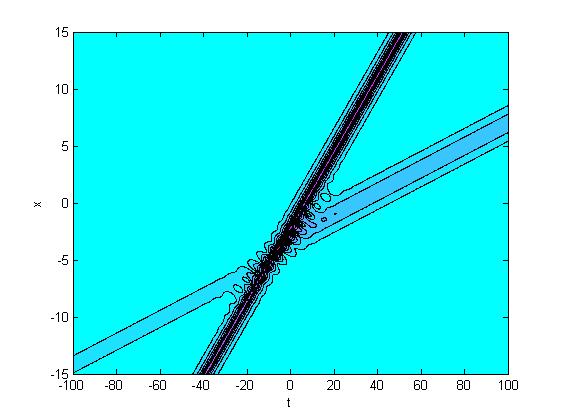}\caption*{(b2)}
\end{minipage}
\caption{\small{(a1)-(a2) describe inelastic collision and energy exchange between two bright-bright soliton solutions to
coupled focusing-defocusing complex short pulse equation. (b1)-(b2) are contour plots.}}
\label{fig2}
\end{figure}
 \begin{figure}[!h]
\centering
\begin{minipage}{0.3\linewidth}\centering
\includegraphics[width=1\textwidth]{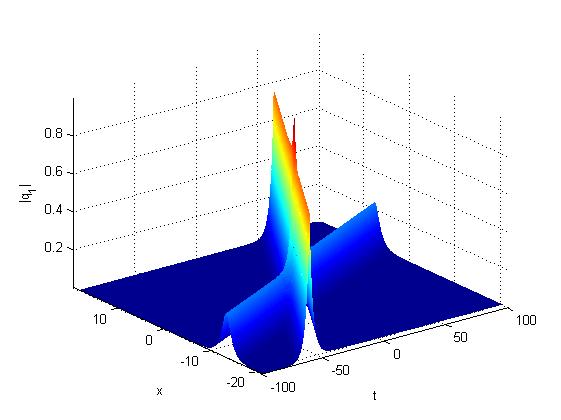}\caption*{(a1)}
\end{minipage}
\begin{minipage}{0.3\linewidth}\centering
\includegraphics[width=1\textwidth]{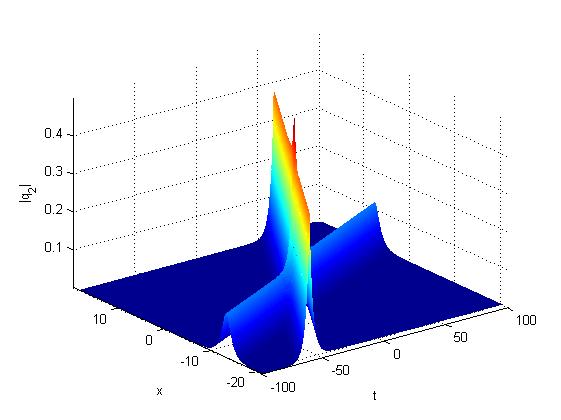}\caption*{(a2)}
\end{minipage}\\
\begin{minipage}{0.3\linewidth}\centering
\includegraphics[width=1\textwidth]{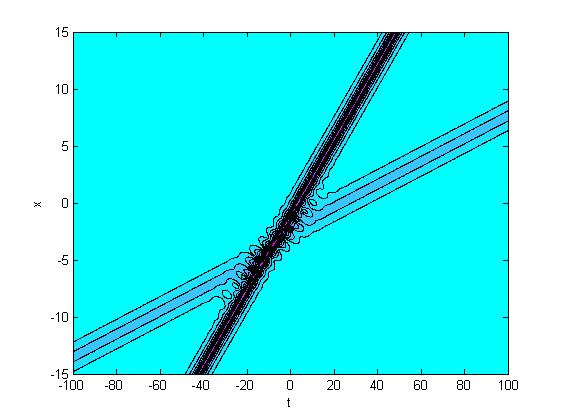}\caption*{(b1)}
\end{minipage}
\begin{minipage}{0.3\linewidth}\centering
\includegraphics[width=1\textwidth]{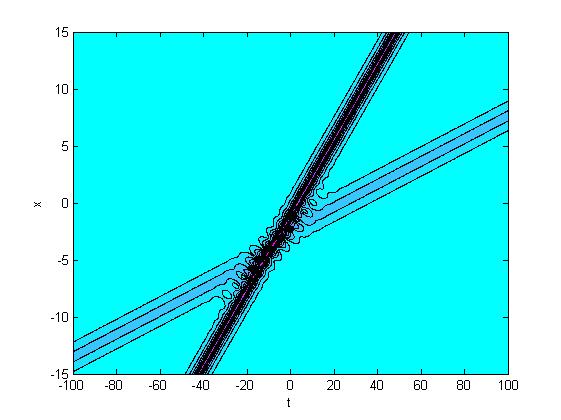}\caption*{(b2)}
\end{minipage}
\caption{\small{(a1)-(a2) describe elastic collision in coupled focusing-defocusing complex short pulse equation. (b1)-(b2) are contour plots.}}
\label{fig3}
\end{figure}
\begin{figure}[!h]
\centering
\begin{minipage}{0.3\linewidth}\centering
\includegraphics[width=1\textwidth]{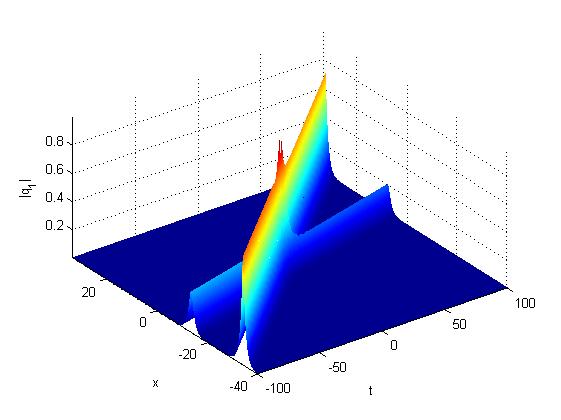}\caption*{(a1)}
\end{minipage}
\begin{minipage}{0.3\linewidth}\centering
\includegraphics[width=1\textwidth]{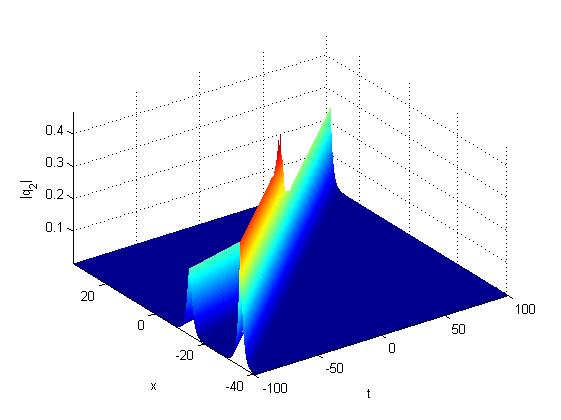}\caption*{(a2)}
\end{minipage}\\
\begin{minipage}{0.3\linewidth}\centering
\includegraphics[width=1\textwidth]{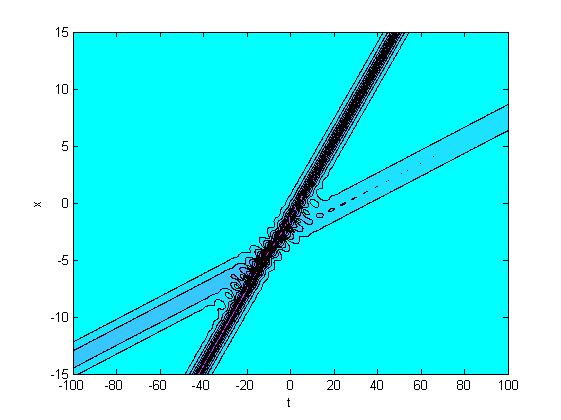}\caption*{(b1)}
\end{minipage}
\begin{minipage}{0.3\linewidth}\centering
\includegraphics[width=1\textwidth]{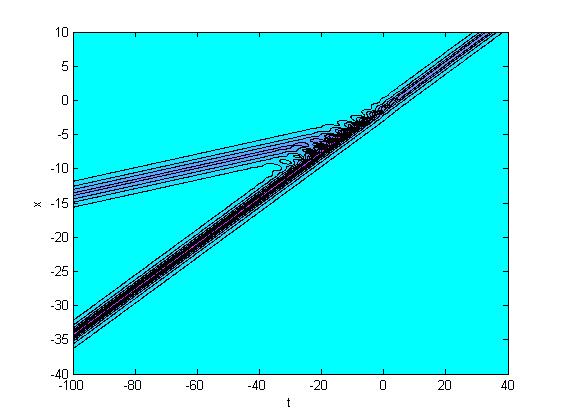}\caption*{(b2)}
\end{minipage}
\caption{\small{(a1)-(a2) describe inelastic collision in coupled focusing-defocusing complex short pulse equation for $p_1=1-\sqrt{2}i, p_2=1+3i, \alpha_{1}=3,\alpha_{2}=2,
\beta_{1}=1,\beta_{2}=0$. (b1)-(b2) are contour plots.}}
\label{fig4}
\end{figure}
\begin{figure}[!h]
\centering
\begin{minipage}{0.3\linewidth}\centering
\includegraphics[width=1\textwidth]{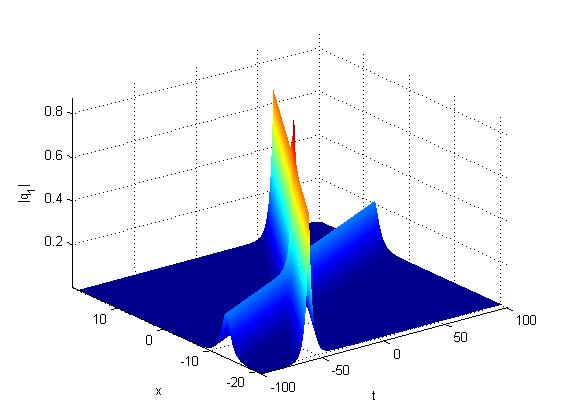}\caption*{(a1)}
\end{minipage}
\begin{minipage}{0.3\linewidth}\centering
\includegraphics[width=1\textwidth]{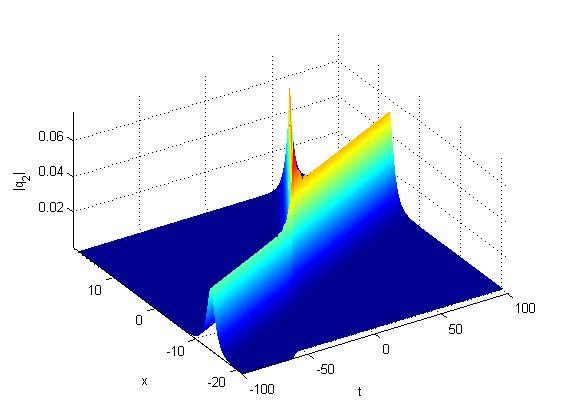}\caption*{(a2)}
\end{minipage}\\
\begin{minipage}{0.3\linewidth}\centering
\includegraphics[width=1\textwidth]{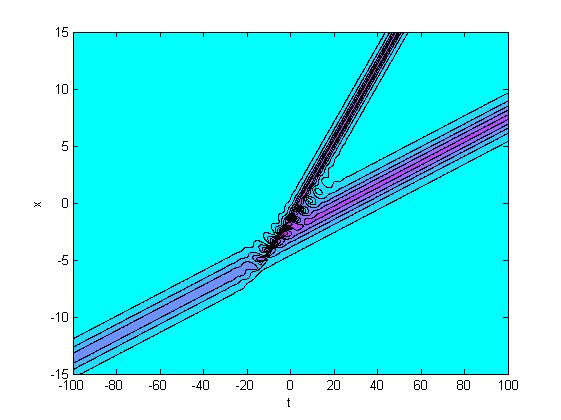}\caption*{(b1)}
\end{minipage}
\begin{minipage}{0.3\linewidth}\centering
\includegraphics[width=1\textwidth]{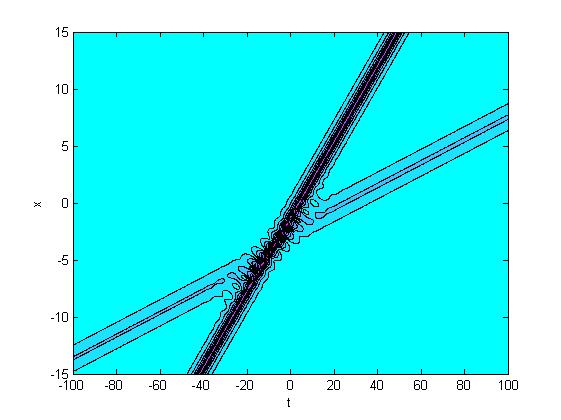}\caption*{(b2)}
\end{minipage}
\caption{\small{(a1)-(a2) describe inelastic collision in coupled focusing-defocusing complex short pulse equation for $p_1=1-\sqrt{2}i, p_2=1+3i,\alpha_{1}=3, \alpha_{2}=2,
\beta_{1}=0, \beta_{2}=1$. (b1)-(b2) are contour plots.}}
\label{fig5}
\end{figure}
\begin{figure}[!h]
\centering
\begin{minipage}{0.3\linewidth}\centering
\includegraphics[width=1\textwidth]{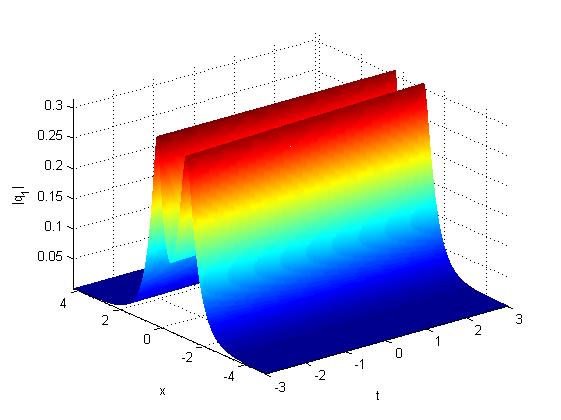}\caption*{(a)}
\end{minipage}
\begin{minipage}{0.3\linewidth}\centering
\includegraphics[width=1\textwidth]{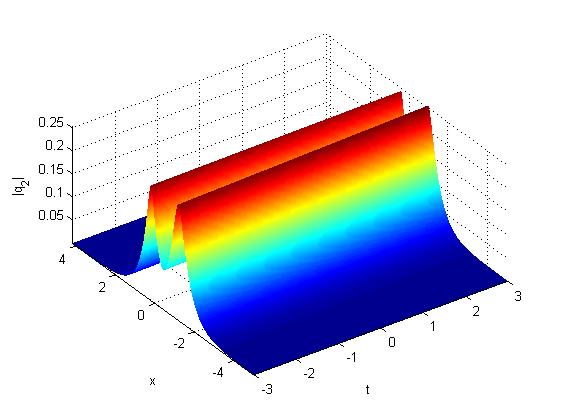}\caption*{(b)}
\end{minipage}
\caption{\small{Parallel two bright-bright solitons in coupled focusing-defocusing complex short pulse equation for $p_1=3+4i, p_2=2+\sqrt{21}i,\alpha_{1}=2, \alpha_{2}=3,
\beta_{1}=1.6, \beta_{2}=2.5$.}}
\label{fig6}
\end{figure}
where
\begin{align}\label{21}  \notag
\eta_{\rm\scriptscriptstyle 1R}&=p_{\rm\scriptscriptstyle 1R}y+\frac{p_{\rm\scriptscriptstyle 1R}}{|p_{1}|^{2}}s+\eta_{\rm\scriptscriptstyle 1,0R},~~~\eta_{\rm\scriptscriptstyle 1I}=p_{\rm\scriptscriptstyle 1I}y-\frac{p_{\rm\scriptscriptstyle 1I}}{|p_{1}|^{2}}s+\eta_{\rm\scriptscriptstyle 1,0I},\\
A_{1}&=\frac{\alpha_{1}}{\sqrt{|\alpha_{1}|^{2}-|\beta_{1}|^{2}}}, A_{2}=\frac{\beta_{1}}{\sqrt{|\alpha_{1}|^{2}-|\beta_{1}|^{2}}},
\eta_{10}=\ln\frac{\sqrt{|\alpha_{1}|^{2}-|\beta_{1}|^{2}}|p_{1}|^{2}}{4|p_{\rm\scriptscriptstyle 1R}|}.
\end{align}
Eq. \eqref{20} is an envelope bright soliton with the amplitude
$2|A_{j}|p_{\rm\scriptscriptstyle 1R}/|p_{1}|^{2}~(j=1,2)$, velocity $-\frac{1}{|p_{1}|^{2}}$ and phase $\eta_{\rm\scriptscriptstyle 10}$.
Let us analyze the property for the one-soliton solution. Notice that
\begin{equation}\label{22}
\frac{\partial x}{\partial y}=1-\frac{2p_{\rm\scriptscriptstyle 1R}^{2}}{|p_{1}|^2}\textup{sech}^{2}(\eta_{\rm\scriptscriptstyle 1R}+\eta_{10}),
\end{equation}
We have $\frac{\partial x}{\partial y}\rightarrow 1$ as $y\rightarrow\pm\infty$. The term $\frac{\partial x}{\partial y}$ attains a minimum
value$(p_{\rm\scriptscriptstyle 1I}^{2}-p_{\rm\scriptscriptstyle 1R}^{2})/(p_{\rm{\scriptscriptstyle 1I}}^{2}+p_{\rm\scriptscriptstyle 1R}^{2})$
at the peak point of soliton wave.
Since $\frac{\partial |q_{j}|}{\partial x}=\frac{\partial |q_{j}|}
{\partial y}/\frac{\partial x}{\partial y}$, we classify
the one-soliton solution as follows:\\
$\bullet$ smooth soliton: when $|p_{\rm\scriptscriptstyle 1R}| < |p_{\rm{\scriptscriptstyle 1I}}|$,
$\frac{\partial x}{\partial y}$ is always positive,
which leads to a smooth envelope soliton. An example
with $p_{1}=0.8+1.3i$ is illustrated in Fig. 1(a).\\
$\bullet$ cuspon soliton: when $|p_{\rm\scriptscriptstyle 1R}|=|p_{\rm{\scriptscriptstyle 1I}}|$,
$\frac{\partial x}{\partial y}$ has a minimum value
of zero at $\eta_{\rm\scriptscriptstyle 1R}+\eta_{10}=0$, which makes the derivative of the envelope
$|q_{j}|$ with respect to $x$ going to infinity at the peak point.
Thus, we have a cusponed envelope soliton, which is illustrated
in Fig. 1(b) with $p_{1}=1+i$.\\
$\bullet$ loop soliton: when $|p_{\rm\scriptscriptstyle 1R}| > |p_{\rm{\scriptscriptstyle 1I}}|$,
the minimum value of $\frac{\partial x}{\partial y}$ becomes negative and
$\frac{\partial x}{\partial y}$ has two zeros at both
sides of the peak of the envelope soliton. This leads to a loop soliton for the envelope
of $q_{j}$. An example is shown in Fig. 1(c) with $p_{1}=0.38-0.21i$.

\textbf{2.2.3 bright-bright soliton solution}

From the N-soliton solution expression \eqref{17}-\eqref{18} of the coupled focusing-defocusing complex short pulse
equation, we obtain two-soliton solution $q_{1}=\frac{g}{f},~q_{2}=\frac{h}{f}$, where
\begin{align}  \notag
f&={\rm Pf}(a_{1},a_{2},a_{3},a_{4},b_{1},b_{2},b_{3},b_{4})\\                                          \label{23}
&=1+e^{\eta_{1}+\bar{\eta}_{1}+\theta_{1\bar{1}}}+
e^{\eta_{1}+\bar{\eta}_{2}+\theta_{1\bar{2}}}+e^{\eta_{2}+\bar{\eta}_{1}+\theta_{2\bar{1}}}
+e^{\eta_{2}+\bar{\eta}_{2}+\theta_{2\bar{2}}}\\ \notag
&+|P_{12}|^{2}|P_{1\bar{2}}|^{2}P_{1\bar{1}}P_{2\bar{2}}(B_{1\bar{1}}B_{2\bar{2}}-B_{2\bar{1}}B_{1\bar{2}})
e^{\eta_{1}+\eta_{2}+\bar{\eta}_{1}+\bar{\eta}_{2}},
\end{align}
\begin{align}  \notag
g&={\rm Pf}(d_{0},\gamma_{1},a_{1},a_{2},a_{3},a_{4},b_{1},b_{2},b_{3},b_{4})\\                         \label{24}
&=\alpha_{1}e^{\eta_{1}}+\alpha_{2}e^{\eta_{2}}
+P_{12}P_{1\bar{1}}P_{2\bar{1}}(\alpha_{2}B_{1\bar{1}}-\alpha_{1}B_{2\bar{1}})e^{\eta_{1}+\eta_{2}+\bar{\eta}_{1}}\\  \notag
&+P_{12}P_{1\bar{2}}P_{2\bar{2}}(\alpha_{2}B_{1\bar{2}}-\alpha_{1}B_{2\bar{2}})e^{\eta_{1}+\eta_{2}+\bar{\eta}_{2}},
\end{align}
\begin{align}\label{25}   \notag
h&={\rm Pf}(d_{0},\gamma_{2},a_{1},a_{2},a_{3},a_{4},b_{1},b_{2},b_{3},b_{4})\\
&=\beta_{1}e^{\eta_{1}}+\beta_{2}e^{\eta_{2}}
+P_{12}P_{1\bar{1}}P_{2\bar{1}}(\beta_{2}B_{1\bar{1}}-\beta_{1}B_{2\bar{1}})e^{\eta_{1}+\eta_{2}+\bar{\eta}_{1}}\\  \notag
&+P_{12}P_{1\bar{2}}P_{2\bar{2}}(\beta_{2}B_{1\bar{2}}-\beta_{1}B_{2\bar{2}})e^{\eta_{1}+\eta_{2}+\bar{\eta}_{2}},
\end{align}
where
\begin{equation}\label{26}
P_{ij}=\frac{p_{i}-p_{j}}{p_{i}+p_{j}},
P_{i\bar{j}}=\frac{p_{i}-\bar{p}_{j}}{p_{i}+\bar{p}_{j}},
B_{i\bar{j}}=\frac{\alpha_{i}\bar{\alpha}_{j}-\beta_{i}\bar{\beta}_{j}}{4(p_{i}^{-2}-\bar{p}_{j}^{-2})},
e^{\theta_{i\bar{j}}}=\frac{\alpha_{i}\bar{\alpha}_{j}-\beta_{i}\bar{\beta}_{j}}{4(p_{i}^{-1}+\bar{p}_{j}^{-1})^{2}}.
\end{equation}
 \textbf{Remark 1}: In order to obtain bright-bright solitons, the conditions $|\alpha_{j}|>|\beta_{j}| (j=1,2)$ still need to be satisfied.
From the forward analysis, we know that the condition
$|p_{\rm\scriptscriptstyle jR}| < |p_{\rm\scriptscriptstyle jI}|$,
$|p_{\rm\scriptscriptstyle jR}| = |p_{\rm\scriptscriptstyle jI}|$,
$|p_{\rm\scriptscriptstyle jR}|>|p_{\rm\scriptscriptstyle jI}|$ leads
 to the smooth soliton, the cuspon soliton and the loop soliton, respectively.
Next, we investigate the asymptotic behavior of the bright-bright soliton
solutions \eqref{23}-\eqref{25} under the reciprocal transformation\eqref{tran1}.
We assume $p_{\rm\scriptscriptstyle 1R},~ p_{\rm\scriptscriptstyle 2R}>0$, $|p_{2}|>|p_{1}|$
without loss of generality. We discuss the following two cases:
(i) when the wave-$\eta_{\rm\scriptscriptstyle 1R}$ is fixed, $\eta_{\rm\scriptscriptstyle 2R}$ can be written as
$\eta_{\rm\scriptscriptstyle 2R}=\frac{p_{\rm\scriptscriptstyle 2R}}{p_{\rm\scriptscriptstyle 1R}}\eta_{\rm\scriptscriptstyle 1R}+
p_{\rm\scriptscriptstyle 2R}(\frac{1}{|p_2|^2}-\frac{1}{|p_1|^2})s$.
When $t\rightarrow \pm\infty$, $\eta_{\rm\scriptscriptstyle 2R}\rightarrow \pm\infty$  for soliton 1.
(ii) the wave-$\eta_{\rm\scriptscriptstyle 2R}$ is fixed, $\eta_{\rm\scriptscriptstyle 1R}$ can be written as
$\eta_{\rm\scriptscriptstyle 1R}=\frac{p_{\rm\scriptscriptstyle 1R}}{p_{\rm\scriptscriptstyle 2R}}\eta_{\rm\scriptscriptstyle 2R}+
p_{\rm\scriptscriptstyle 1R}(\frac{1}{|p_1|^2}-\frac{1}{|p_2|^2})s$.
When $t\rightarrow \pm\infty$,  $\eta_{\scriptscriptstyle 1R}\rightarrow \mp\infty$ for soliton 2.
This leads to the following asymptotic forms for two-soliton solution.\\
(i) Before collision $(t\rightarrow -\infty)$:\\
Soliton 1(the wave-$\eta_{\rm\scriptscriptstyle 1R}$ is fixed, $\eta_{\rm\scriptscriptstyle 2R}\rightarrow -\infty$),
\begin{align}\label{27}
\left(
  \begin{array}{c}
    q_{1} \\
    q_{2} \\
  \end{array}
\right)&\rightarrow\left(
  \begin{array}{c}
    \alpha_{1} \\
    \beta_{1} \\
  \end{array}
\right)\frac{e^{\eta_{1}}}{1+e^{\eta_{1}+\eta^{*}_{1}+\theta_{1\bar{1}}}}                     \\
&=
\left(
  \begin{array}{c}
    A_{1}^{1-} \\
    A_{2}^{1-} \\
  \end{array}
\right)\frac{2p_{\scriptscriptstyle 1R}}{|p_{1}|^{2}}e^{i\eta_{\scriptscriptstyle 1I}}\textup{sech}(\eta_{\scriptscriptstyle 1R}+\frac{\theta_{1\bar{1}}}{2}),  \notag
\end{align}
where
\begin{align}\label{28}
\left(
  \begin{array}{c}
    A_{1}^{1-} \\
    A_{2}^{1-} \\
  \end{array}
\right)=\left(
  \begin{array}{c}
    \alpha_{1} \\
    \beta_{1} \\
  \end{array}
\right)\frac{1}{\sqrt{|\alpha_{1}|^2-|\beta_1|^2}}.
\end{align}
Soliton 2(the wave-$\eta_{\rm\scriptscriptstyle 2R}$ is fixed, $\eta_{\rm\scriptscriptstyle 1R}\rightarrow +\infty$),
\begin{align}\label{29}
\left(
  \begin{array}{c}
    q_{1} \\
    q_{2} \\
  \end{array}
\right)\rightarrow\left(
  \begin{array}{c}
    A_{1}^{2-} \\
    A_{2}^{2-} \\
  \end{array}
\right)\frac{2p_{\scriptscriptstyle 2R}}{|p_{2}|^{2}}e^{i\eta_{\scriptscriptstyle 2I}}\textup{sech}(\eta_{\scriptscriptstyle 2R}+\frac{\theta_{1\bar{1}2\bar{2}}-\theta_{1\bar{1}}}{2}),
\end{align}
where
\begin{align}\label{30}
\left(
  \begin{array}{c}
    A_{1}^{2-} \\
    A_{2}^{2-} \\
  \end{array}
\right)=\left(
  \begin{array}{c}
   e^{\theta_{12\bar{1}}^{(1)}} \\
   e^{\theta_{12\bar{1}}^{(2)}} \\
  \end{array}
\right)\frac{e^{-(\theta_{1\bar{1}2\bar{2}}+\theta_{1\bar{1}}-\theta_{2\bar{2}})/2}}{\sqrt{|\alpha_{2}|^2-|\beta_2|^2}},
\end{align}
with
\begin{align}\label{31}  \notag
\theta_{12\bar{1}}^{(1)}&=P_{12}P_{1\bar{1}}P_{2\bar{1}}(\alpha_{2}B_{1\bar{1}}-\alpha_{1}B_{2\bar{1}}),\\ \notag
\theta_{12\bar{1}}^{(2)}&=P_{12}P_{1\bar{1}}P_{2\bar{1}}(\beta_{2}B_{1\bar{1}}-\beta_{1}B_{2\bar{1}}),\\   \notag
\theta_{1\bar{1}2\bar{2}}&=P_{12}P_{1\bar{2}}P_{2\bar{2}}(\alpha_{2}B_{1\bar{2}}-\alpha_{1}B_{2\bar{2}}).
\end{align}

(ii) After collision $(t\rightarrow +\infty)$:\\
Soliton 1(the wave-$\eta_{\rm\scriptscriptstyle 1R}$ is fixed, $\eta_{\scriptscriptstyle 2R}\rightarrow +\infty$),
\begin{equation}\label{32}
\left(
  \begin{array}{c}
    q_{1} \\
    q_{2} \\
  \end{array}
\right)\rightarrow\left(
  \begin{array}{c}
    A_{1}^{1+} \\
    A_{2}^{1+} \\
  \end{array}
\right)\frac{2p_{\scriptscriptstyle 1R}}{|p_{1}|^{2}}e^{i\eta_{\scriptscriptstyle 1I}}\textup{sech}(\eta_{\scriptscriptstyle 1R}+\frac{\theta_{1\bar{1}2\bar{2}}-\theta_{2\bar{2}}}{2}),
\end{equation}
where
\begin{equation}\label{33}
\left(
  \begin{array}{c}
    A_{1}^{1+} \\
    A_{2}^{1+} \\
  \end{array}
\right)=\left(
  \begin{array}{c}
   e^{\theta_{12\bar{2}}^{(1)}} \\
   e^{\theta_{12\bar{2}}^{(2)}} \\
  \end{array}
\right)\frac{e^{-(\theta_{1\bar{1}2\bar{2}}+\theta_{2\bar{2}}-\theta_{1\bar{1}})/2}}{\sqrt{|\alpha_{1}|^2-|\beta_1|^2}},
\end{equation}
with
\begin{align}\label{34}  \notag
\theta_{12\bar{2}}^{(1)}&=P_{12}P_{1\bar{2}}P_{2\bar{2}}(\alpha_{2}B_{1\bar{2}}-\alpha_{1}B_{2\bar{2}}),\\ \notag
\theta_{12\bar{2}}^{(2)}&=P_{12}P_{1\bar{2}}P_{2\bar{2}}(\beta_{2}B_{1\bar{2}}-\beta_{1}B_{2\bar{2}}).
\end{align}
Soliton 2:(the wave-$\eta_{\rm\scriptscriptstyle 2R}$ is fixed, $\eta_{\scriptscriptstyle 1R}\rightarrow -\infty$),
\begin{equation}\label{35}
\left(
  \begin{array}{c}
    q_{1} \\
    q_{2} \\
  \end{array}
\right)
\rightarrow
\left(
  \begin{array}{c}
    A_{1}^{2+} \\
    A_{2}^{2+} \\
  \end{array}
\right)\frac{2p_{\scriptscriptstyle 2R}}{|p_{2}|^{2}}e^{i\eta_{\scriptscriptstyle 2I}}\textup{sech}(\eta_{\scriptscriptstyle 2R}+\frac{\theta_{2\bar{2}}}{2}),  \notag
\end{equation}
where
\begin{equation}\label{36}
\left(
  \begin{array}{c}
    A_{1}^{2+} \\
    A_{2}^{2+} \\
  \end{array}
\right)=\left(
  \begin{array}{c}
    \alpha_{2} \\
    \beta_{2} \\
  \end{array}
\right)\frac{1}{\sqrt{|\alpha_{2}|^2-|\beta_2|^2}}.
\end{equation}
\begin{table}[htbp]
 \caption{Physical quantities of bright-bright soliton $S_{1}$ and soliton $S_{2}$ before and after interactions.}
 \begin{tabular}{l}
  \toprule
  Soliton~~~~~~~~~Amplitude/depth~~~~~~~~~Velocity~~~~~~~~~~~Soliton~~~~~~~~~Amplitude/depth~~~~~~~~~~~Velocity \\
     \midrule
  $S_{1}^{1-}$~~~~~~~~~~~~$\frac{2\alpha_{1}p_{\rm\scriptscriptstyle 1R}}{|p_{1}|^2\sqrt{|\alpha_{1}|^2-|\beta_{1}|^2}}$~~~~~~~~~~~~~$-\frac{1}{|p_{1}|^{2}}$
  ~~~~~~~~~~~~$S_{1}^{1+}$~~~~~~~~~~~~~$\frac{2p_{\rm\scriptscriptstyle 1R}e^{\theta^{(1)}_{12\bar{2}}-(\theta_{1\bar{1}2\bar{2}}+\theta_{2\bar{2}}-\theta_{1\bar{1}})/2}}{|p_{1}|^2\sqrt{|\alpha_{1}|^2-|\beta_{1}|^2}}$
  ~~~~~~~~$-\frac{1}{|p_{1}|^{2}}$ \\

  $S_{2}^{1-}$~~~~~~~~~~~~~$\frac{2\beta_{1}p_{\rm\scriptscriptstyle 1R}}{|p_{1}|^2\sqrt{|\alpha_{1}|^2-|\beta_{1}|^2}}$~~~~~~~~~~~~$-\frac{1}{|p_{1}|^{2}}$
  ~~~~~~~~~~~~$S_{2}^{1+}$~~~~~~~~~~~~~~$\frac{2p_{\rm\scriptscriptstyle 1R}e^{\theta^{(2)}_{12\bar{2}}-(\theta_{1\bar{1}2\bar{2}}+\theta_{2\bar{2}}-\theta_{1\bar{1}})/2}}{|p_{1}|^2\sqrt{|\alpha_{1}|^2-|\beta_{1}|^2}}$
  ~~~~~~~$-\frac{1}{|p_{1}|^{2}}$ \\

  $S_{1}^{2-}$~~~~~~~$\frac{2p_{\rm\scriptscriptstyle 2R}e^{\theta^{(1)}_{12\bar{1}}-(\theta_{1\bar{1}2\bar{2}}+\theta_{1\bar{1}}-\theta_{2\bar{2}})/2}}{|p_{2}|^2
  \sqrt{|\alpha_{2}|^2-|\beta_{2}|^2}}$~~~~~~~$-\frac{1}{|p_{2}|^{2}}$~~~~~~~~~~~~~$S_{1}^{2+}$
  ~~~~~~~~~~~~~$\frac{2\alpha_{2}p_{\rm\scriptscriptstyle 2R}}{|p_{2}|^2\sqrt{|\alpha_{2}|^2-|\beta_{2}|^2}}$
  ~~~~~~~~~~~~~~$-\frac{1}{|p_{2}|^{2}}$ \\

  $S_{2}^{2-}$~~~~~~~$\frac{2p_{\rm\scriptscriptstyle 2R}e^{\theta^{(2)}_{12\bar{1}}-(\theta_{1\bar{1}2\bar{2}}+\theta_{1\bar{1}}-\theta_{2\bar{2}})/2}}{|p_{2}|^2
  \sqrt{|\alpha_{2}|^2-|\beta_{2}|^2}}$~~~~~~~$-\frac{1}{|p_{2}|^{2}}$~~~~~~~~~~~~~$S_{2}^{2+}$
  ~~~~~~~~~~~~~$\frac{2\beta_{2}p_{\rm\scriptscriptstyle 2R}}{|p_{2}|^2\sqrt{|\alpha_{2}|^2-|\beta_{2}|^2}}$
  ~~~~~~~~~~~~~~$-\frac{1}{|p_{2}|^{2}}$ \\
  \bottomrule
 \end{tabular}
\end{table}

Similar to the analysis of the coupled focusing-focusing complex short pulse equation \cite{20} and the coupled focusing-focusing NLS equation \cite{26,27,28}, we introduce the transition matrix $T=[T_{j}^{l}]$ defined by $A_{j}^{l+}=T_{j}^{l}A_{j}^{l-}~(j,l=1,2)$,
and set $\gamma_{1}=\frac{\alpha_{2}}{\alpha_{1}}$,~$\gamma_{2}=\frac{\beta_{2}}{\beta_{1}}$.
Then we have
\begin{align*}
T_{j}^{1}&=-\frac{P_{12}P_{1\bar{2}}}{|P_{12}P
_{1\bar{2}}|}\frac{1}{\sqrt{1-\lambda_{1}\lambda_{2}}}\left(1-\lambda_{2}\gamma_{j}\right),\\
T_{j}^{2}&=\left(\frac{P_{12}P_{2\bar{1}}}{|P_{12}P_{1\bar{2}}|}\right)^{-1}
\sqrt{1-\lambda_{1}\lambda_{2}}\left(1-\lambda_{1}\gamma_{j}^{-1}\right)^{-1}~(j=1,2),
\end{align*}
where $\lambda_{1}=B_{2\bar{1}}/B_{1\bar{1}}$, $\lambda_{2}=B_{1\bar{2}}/B_{2\bar{2}}$.

%thus we have
%\begin{align*}
%T_{j}^{1}&=\left(\frac{P_{12}P_{1\bar{2}}}{\bar{P}_{12}\bar{P}
%_{1\bar{2}}}\right)^{1/2}(\sqrt{1-\lambda_{1}\lambda_{2}})^{-1}\left(1-\lambda_{2}\frac{\alpha_{2}^{(j)}}{\alpha_{1}^{(j)}}\right)\\
%T_{j}^{2}&=\left(\frac{\bar{P}_{12}P_{1\bar{2}}}{P_{12}\bar{P}
%_{1\bar{2}}}\right)^{1/2}\sqrt{1-\lambda_{1}\lambda_{2}}\left(1-\lambda_{1}\frac{\beta_{1}^{(j)}}{\beta_{2}^{(j)}}\right)^{-1}~(j=1,2),
%\end{align*}
%where $\lambda_{1}=B_{2\bar{1}}/B_{1\bar{1}}$, $\lambda_{2}=B_{1\bar{2}}/B_{2\bar{2}}$.

From the above analysis, we can see that there exist an exchange of energies between the
two solitons after the collision. An example
is shown in Fig.2 for the parameters $p_1=1-\sqrt{2}i,~p_2=1+3i,~
\alpha_{1}=5,~\alpha_{2}=7, \beta_{1}=1,~\beta_{2}=2$.
However, only for the special case $\alpha_{1}/\alpha_{2}=\beta_{1}/\beta_{2}$
there is no energy exchange between two components of solitons
after the collision. An example is shown in Fig.3 for the parameters
$p_1=1-\sqrt{2}i,~p_2=1+3i,~\alpha_{1}=2,~\alpha_{2}=4,~\beta_{1}=1,~\beta_{2}=2$.

Fig.4 and Fig.5 show the energy centralization and distribution between the
two solitons $q_1$ and $q_2$ in the process of collision respectively.
We just change the parameters
in previous two examples as $\beta_{1}=0,~\beta_{2}=1$ for Fig.4 and
$\beta_{1}=1,~\beta_{2}=0$ for Fig.5. When $|p_{1}|=|p_{2}|$, the parallel solitons will occur (see Fig. 6).

\section{Bright-dark soliton and dark-dark soliton}
\subsection{bright-dark soliton solution}

\begin{figure}[!h]
\centering
\begin{minipage}{0.3\linewidth}\centering
\includegraphics[width=1\textwidth]{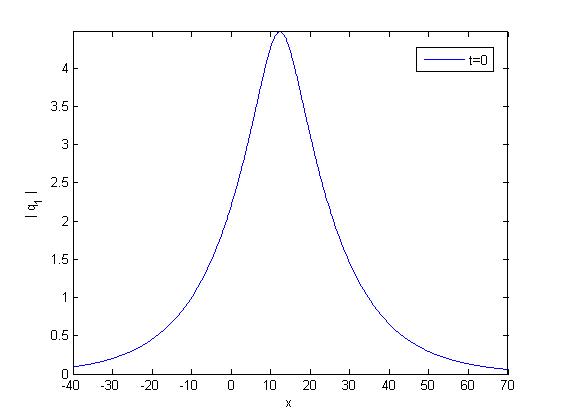}\caption*{(a1)}
\end{minipage}
\begin{minipage}{0.3\linewidth}\centering
\includegraphics[width=1\textwidth]{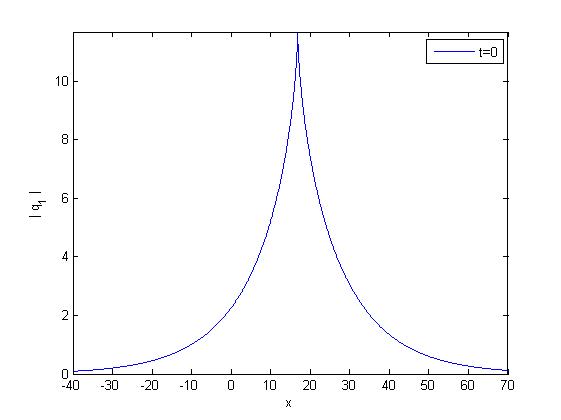}\caption*{(a2)}
\end{minipage}
\begin{minipage}{0.3\linewidth}\centering
\includegraphics[width=1\textwidth]{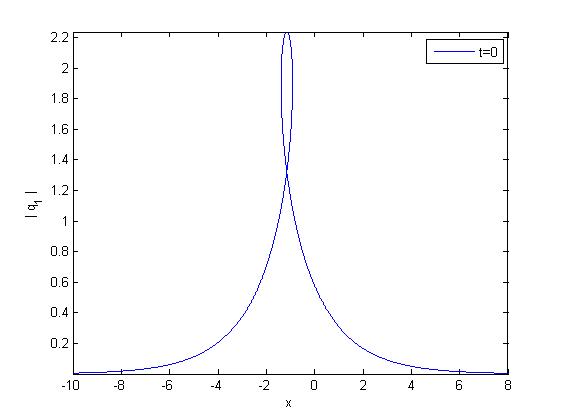}\caption*{(a3)}
\end{minipage}\\
\begin{minipage}{0.3\linewidth}\centering
\includegraphics[width=1\textwidth]{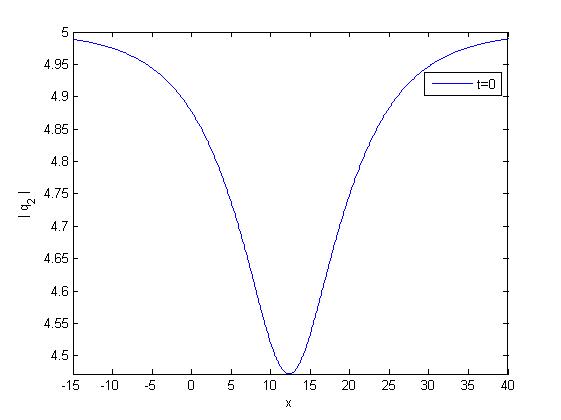}\caption*{(b1)}
\end{minipage}
\begin{minipage}{0.3\linewidth}\centering
\includegraphics[width=1\textwidth]{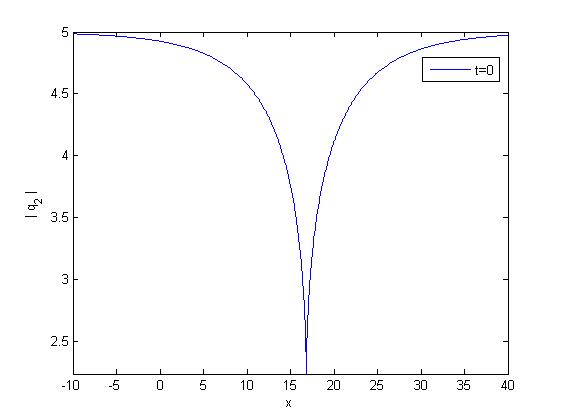}\caption*{(b2)}
\end{minipage}
\begin{minipage}{0.3\linewidth}\centering
\includegraphics[width=1\textwidth]{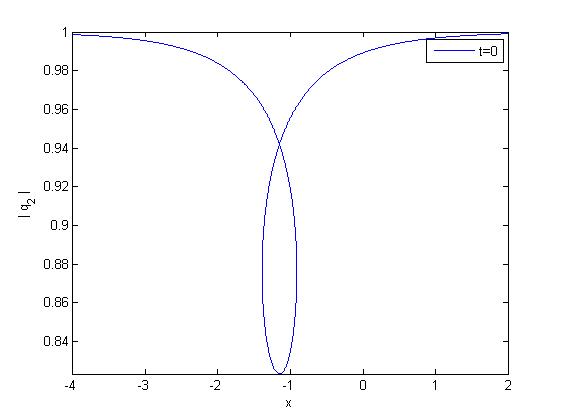}\caption*{(b3)}
\end{minipage}
\caption{\small{One bright and dark soliton solution of $q_{1}$ and $q_{2}$ respectively to the coupled
focusing-defocusing complex short pulse equation with parameters $k=1,
\tau_{1}=5, a_{1}=1+2i, b_1=0$. (a1)-(b1) smooth soliton with
$p_1=1+2i$. (a2)-(b2) cuspon soliton with $p_1=1+i$, (a3)-(b3) loop soliton with $\tau_{1}=1, p_1=0.3-0.2i$.}}
\label{fig7}
\end{figure}

In order to obtain bright-dark soliton, we bilinearize the equation \eqref{9} as
\begin{align}  \notag
&D_{s}D_{y}f\cdot g=f\cdot g, \\ \label{39}
&(D_{s}D_{y}-\lambda)f\cdot h=0, \\
&(D_{s}^{2}-\lambda)f\cdot f=\frac{1}{2}(|g|^{2}-|h|^{2}),\notag
\end{align}
where $\lambda$ is a constant to be determined.
By the similar procedure of obtaining Eqs.\eqref{10}-\eqref{16},
one can check that the bilinear form \eqref{39} can convert
to the Eq.\eqref{9} by the following hodograph transformation
\begin{equation}\label{tran2}
x=\lambda (y+s)-2(lnf)_{s},~~t=-s.
\end{equation}

To get the bright-dark soliton solution, we assume $g=\chi g_{1},~
h=h_{0}(1+\chi^{2} h_{2})$, and $f=1+\chi^{2}f_{2}$.
Substituting them to the bilinear equation \eqref{39}
and comparing the coefficient of the same power of $\chi^{0},~\chi^{1},$
we get
\begin{equation}
h_0=\tau_1e^{i\psi},~~g_1=a_1e^{\eta_{1}}, \notag
\end{equation}
\begin{figure}[!h]
\centering
\begin{minipage}{0.3\linewidth}\centering
\includegraphics[width=1\textwidth]{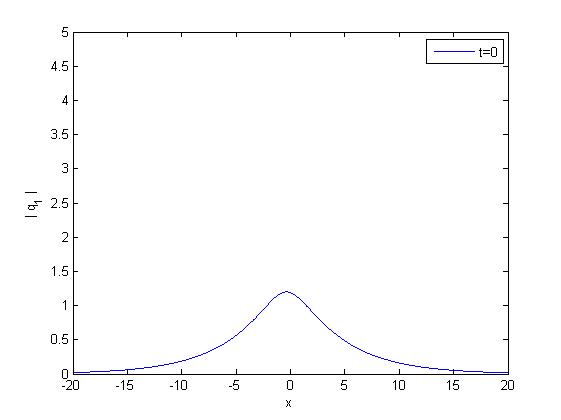}\caption*{(a1)}
\end{minipage}
\begin{minipage}{0.3\linewidth}\centering
\includegraphics[width=1\textwidth]{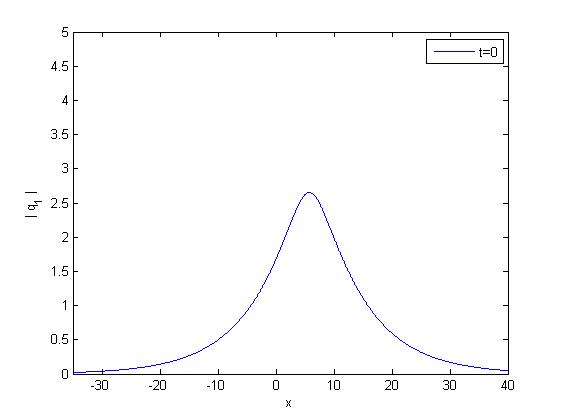}\caption*{(a2)}
\end{minipage}
\begin{minipage}{0.3\linewidth}\centering
\includegraphics[width=1\textwidth]{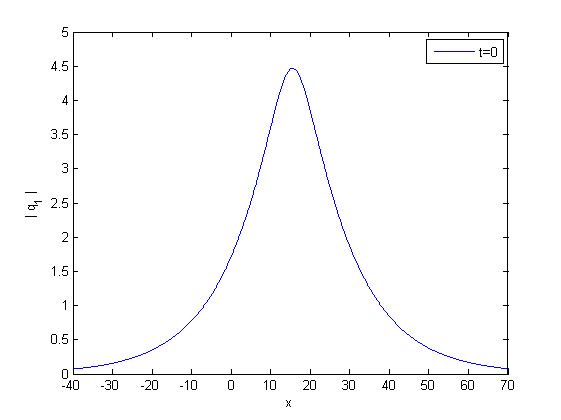}\caption*{(a3)}
\end{minipage}\\
\begin{minipage}{0.3\linewidth}\centering
\includegraphics[width=1\textwidth]{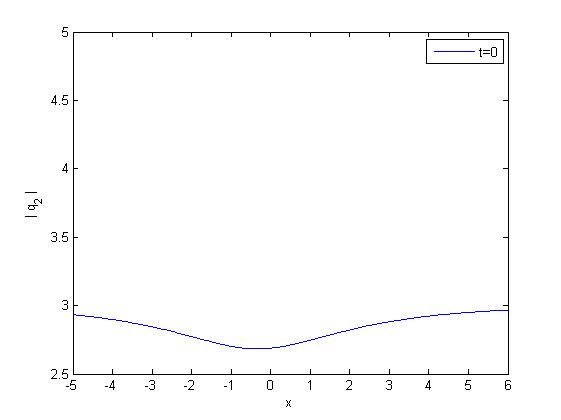}\caption*{(b1)}
\end{minipage}
\begin{minipage}{0.3\linewidth}\centering
\includegraphics[width=1\textwidth]{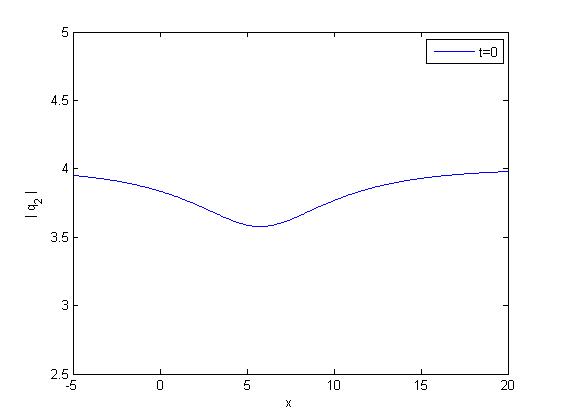}\caption*{(b2)}
\end{minipage}
\begin{minipage}{0.3\linewidth}\centering
\includegraphics[width=1\textwidth]{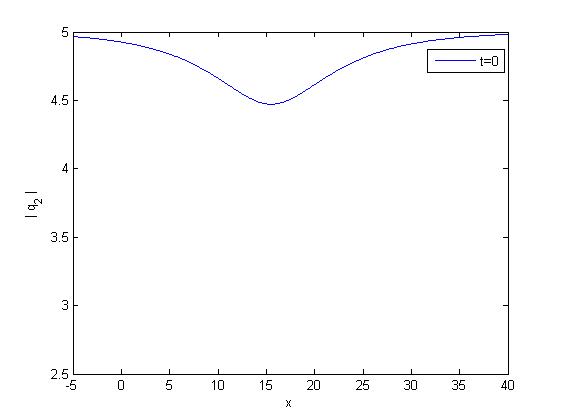}\caption*{(b3)}
\end{minipage}\\
\caption{\small{ Intensity plots of one-soliton of the coupled
focusing-defocusing complex short pulse equation for different values of the background parameter $\tau_{1}$ for the
fixed values of $k=1, a_{1}=1+\sqrt{2}i, p_{1}=1+2i, b_{1}=0$. Note that the intensity of the bright soliton increases as the depth of the dark soliton decreases. The other parameters are
chosen as $b_{1}=0$, $\tau_{1}=3,4,5$ respectively.}}
\label{fig8}
\end{figure}
where $\psi=ky-\frac{\lambda}{k}s,~
\eta_{1}=p_{1}y+\frac{\lambda}{p_{1}}s+b_{1},~\lambda=\frac{1}{2}|\tau_{1}|^{2}$, and here $k$ is an
arbitrary real parameter, $ \tau_{1},~a_{1},~p_{1},~b_{1}$ are arbitrary complex parameters.
 Substituting the form of $h_0$ into \eqref{39} then the bilinear equations change into
\begin{align}\label{40}
&D_{s}D_{y}f\cdot g=f\cdot g, \\   \notag
&(D_{s}D_{y}-ikD_{s}+\frac{i\lambda}{k}D_{y})f\cdot \tilde{h}=0, \\
&(D_{s}^{2}-\lambda)f\cdot f=\frac{1}{2}(|g|^{2}-\tau_{1}^{2}|\tilde{h}|^{2}),               \notag
\end{align}
where $\tilde{h}=1+\chi^{2} h_{2}$.
We set $h_{2}=A_{1\bar{1}}e^{\eta_{1}+\eta^{*}_{1}}, f_{2}=B_{1\bar{1}}e^{\eta_{1}+\eta^{*}_{1}}$,
$\eta_{1}=\eta_{\rm\scriptscriptstyle1R}+i\eta_{\rm\scriptscriptstyle 1I}$ and
$p_{1}=p_{\rm\scriptscriptstyle 1R}+ip_{\rm\scriptscriptstyle 1I}$.
This yields the bright-dark soliton solution
\begin{align}\notag
q_{1}&=\frac{a_{1}e^{\eta_{1}}}{1+B_{1\bar{1}}e^{\eta_{1}+\eta^{*}_{1}}}=\frac{a_{1}}{2}e^{i\eta_{\rm\scriptscriptstyle 1I}-\eta_{10}}\textup{sech}(\eta_{\rm\scriptscriptstyle 1R}+\eta_{10}),\\  \label{41}
q_{2}&=\tau_{1}e^{i\psi}\frac{1+A_{1\bar{1}}e^{\eta_{1}+\eta^{*}_{1}}}{1+B_{1\bar{1}}e^{\eta_{1}+\eta^{*}_{1}}}
=\frac{1}{2}\tau_{1}e^{i\psi}[1+\mu_{1\bar{1}}-(1-\mu_{1\bar{1}})\tanh(\eta_{\rm\scriptscriptstyle 1R}+\eta_{10})],\\  \notag
x&=\lambda(y+s-2\frac{p_{\rm\scriptscriptstyle 1R}}{|p_{1}|^{2}}(\tanh(\eta_{\rm\scriptscriptstyle 1R}+\eta_{10})+1)),~~~t=-s,
\end{align}
where
\begin{align}  \notag
\eta_{\rm\scriptscriptstyle 1R}&=p_{\rm\scriptscriptstyle 1R}y+\frac{\lambda p_{\rm\scriptscriptstyle 1R}}{|p_{1}|^{2}}s+b_{\rm1\scriptscriptstyle R},~~~\eta_{\rm\scriptscriptstyle 1I}=p_{\rm\scriptscriptstyle 1I}y-\frac{\lambda p_{\rm\scriptscriptstyle 1I}}{|p_{1}|^{2}}s+b_{\rm1\scriptscriptstyle I},
~~\eta_{10}=\frac{1}{2}\ln B_{1\bar{1}},\\   \notag
B_{1\bar{1}}&=\frac{|a_{1}|^{2}|p_{1}|^{4}(k^{2}+p_{1}^{2})(k^{2}+p_{1}^{*2})}
{4\lambda(p_{1}+p_{1}^{*})^{2}[\lambda k^{2}(k^{2}+p_{1}^{2}+p_{1}^{*2})+|p_{1}|^{4}(\lambda-2k^{2})]},\\ \notag
\mu_{1\bar{1}}&=\frac{(k+ip_{1})(k+ip_{1}^{*})}{(k-ip_{1})(k-ip_{1}^{*})},  A_{1\bar{1}}=\mu_{1\bar{1}}B_{1\bar{1}}.
\end{align}
In order to avoid singularity, we emphasize $B_{1\bar{1}}>0$, which means
\begin{equation}  \label{43}
|\tau_{1}|>\frac{2|kp_{1}|^{2}}{|p_{1}^{2}+k^{2}|}.
\end{equation}
\begin{figure}[!h]
\centering
\begin{minipage}{0.3\linewidth}\centering
\includegraphics[width=1\textwidth]{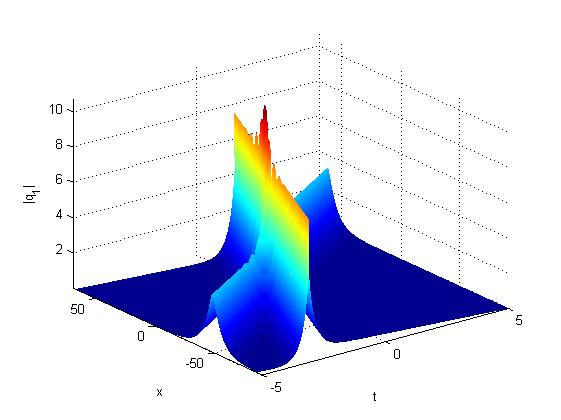}\caption*{(a)}
\end{minipage}
\begin{minipage}{0.3\linewidth}\centering
\includegraphics[width=1\textwidth]{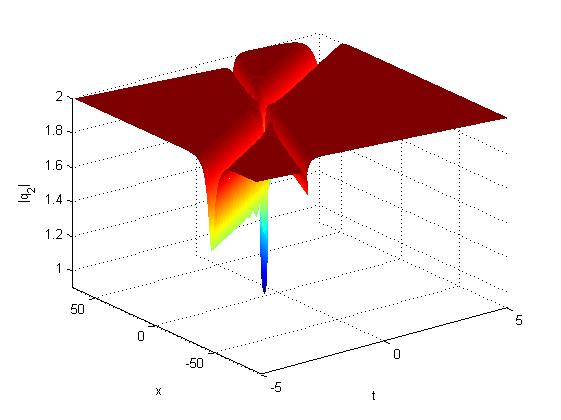}\caption*{(b)}
\end{minipage}\\
\caption{\small{Two bright-dark soliton solution to the coupled focusing-defocusing
complex short pulse equation with parameters $\tau_{1}=2, k=1,
p_{1}=\frac{1}{5}+\frac{1}{4}i, p_{2}=\frac{3}{10}-\frac{1}{2}i,
\alpha_{1}=1, \alpha_{2}=\frac{1}{2}, b_{1}=b_{2}=0.$}}
\label{fig9}
\end{figure}
Notice that
\begin{equation}\label{43}
\frac{\partial x}{\partial y}=\lambda-\frac{2\lambda p_{\scriptscriptstyle 1R}^{2}}{|p_{1}|^2}\textup{sech}^{2}(\eta_{\scriptscriptstyle 1R}+\eta_{10}),
\end{equation}
we have $\frac{\partial x}{\partial y}\rightarrow \lambda$ as $y\rightarrow\pm\infty.$
  Since $\partial |q_{i}|/\partial x=\frac{\partial |q_{i}|/\partial y}{\partial x/\partial y}$, we can also classify
the one-soliton solution as smooth, cuspon and loop soliton respectively when $|p_{\rm\scriptscriptstyle 1R}| < |p_{\rm\scriptscriptstyle 1I}|$, $|p_{\rm\scriptscriptstyle 1R}|=|p_{\rm\scriptscriptstyle 1I}|$ and $|p_{\rm\scriptscriptstyle 1R}| > |p_{\rm\scriptscriptstyle 1I}|$.
An example is shown in Fig. 7. The amplitude of the bright soliton is $\frac{\alpha_{1}}{2\sqrt{B_{1\bar{1}}}}$ with
the velocity $-\frac{\lambda p_{\rm\scriptscriptstyle 1R}}{|P_{1}|^{2}}$ in the (y,s)-coordinate system and $|\tau_{1}(1-\mu_{1\bar{1}})|/2$ is
amplitude of the dark soliton related to the background energy.
We can observe that with the background soliton parameter $|\tau_{1}|$ increasing,
the amplitude of the bright soliton increases simultaneously (see Fig. 8),
which is different from the coupled nonlinear Schr\"{o}dinger equation\cite{25}.

Then the two bright-dark soliton solutions for Eqs.\eqref{4} is $q_{1}=\frac{g}{f},~q_{2}=\frac{h}{f}$, where
\begin{align}\notag
f&=1+B_{1\bar{1}}e^{\eta_{1}+\bar{\eta}_{1}}+B_{1\bar{2}}e^{\eta_{1}+\bar{\eta}_{2}}+B_{2\bar{1}}e^{\eta_{2}+\bar{\eta}_{1}}
+B_{2\bar{2}}e^{\eta_{2}+\bar{\eta}_{2}}\\ \notag
&+|P_{12}|^{2}(P_{1\bar{1}}P_{2\bar{2}}B_{1\bar{2}}B_{2\bar{1}}
-P_{1\bar{2}}P_{\bar{2}1}B_{2\bar{2}}B_{1\bar{2}})e^{\eta_{1}+\eta_{2}+\bar{\eta}_{1}+\bar{\eta}_{2}},\\ \label{44}
g&=a_{1}e^{\eta_{1}}+a_{2}e^{\eta_{2}}
+P_{12}(a_{1}P_{1\bar{1}}B_{2\bar{1}}-a_{2}P_{2\bar{1}}B_{1\bar{1}})e^{\eta_{1}+\eta_{2}+\bar{\eta}_{1}}\\  \notag
&+P_{12}(a_{1}P_{1\bar{2}}B_{2\bar{2}}-a_{2}P_{2\bar{2}}B_{1\bar{2}})e^{\eta_{1}+\eta_{2}+\bar{\eta}_{2}},\\        \notag
h&=\tau_{1}e^{i(ky-\frac{\lambda}{k}s)}(1+A_{1\bar{1}}e^{\eta_{1}+\bar{\eta}_{1}}+A_{1\bar{2}}e^{\eta_{1}
+\bar{\eta}_{2}}+A_{2\bar{1}}e^{\eta_{2}+\bar{\eta}_{1}}+A_{2\bar{2}}e^{\eta_{2}+\bar{\eta}_{2}}\\ \notag
&+\mu_{1\bar{1}}\mu_{2\bar{2}}|P_{12}|^{2}(P_{1\bar{1}}P_{2\bar{2}}B_{1\bar{2}}
B_{2\bar{1}}-P_{1\bar{2}}P_{1\bar{1}}B_{2\bar{2}}B_{1\bar{2}})
e^{\eta_{1}+\eta_{2}+\bar{\eta}_{1}+\bar{\eta}_{2}}),\\   \notag
x&=\lambda (y+s)-2(lnf)_{s},~~t=-s.
\end{align}
where
\begin{equation}\label{45}  \notag
B_{j\bar{l}}=\frac{a_{j}a^{*}_{l}p_{j}^{2}p^{*2}_{l}(k^{2}+p^{2}_{j})(k^{2}+p^{*2}_{l})}
{4\lambda(p_{j}+p^{*}_{l})^{2}(\lambda k^{2}(k^{2}+p_{j}^{2}+p_{l}^{*2})+p_{j}^{2}p_{l}^{*2}(\lambda-2k^{2}))}, \notag
\end{equation}
\begin{equation}\label{46}  \notag
A_{j\bar{l}}=\mu_{j\bar{l}}B_{j\bar{l}},~~P_{ij}=\frac{p_{j}-p_{l}}
{p_{j}+p_{l}},~~P_{j\bar{l}}=\frac{p_{j}-\bar{p}_{l}}{p_{j}+\bar{p}_{l}},\\  \notag
\end{equation}
\begin{equation}\label{47}  \notag
\mu_{j\bar{l}}=\frac{(k+ip_{j})(k+ip^{*}_{l})}{(k-ip_{j})(k-ip^{*}_{l})},~~~
\eta_{j}=p_{j}y+\frac{\lambda}{p_{j}}s+b_{j}~~(j,l=1,2).
\end{equation}
Due to $|\mu_{j\bar{l}}|=1$, we assume $\mu_{j\bar{l}}=e^{i\phi_{jl}}$.
Next we investigate the asymptotic behavior of the two bright-dark soliton solutions.\\
(i) Before collision $(t\rightarrow -\infty)$:\\
Soliton 1:(the wave-$\eta_{\rm\scriptscriptstyle 1R}$ is fixed, $\eta_{\scriptscriptstyle 2R}\rightarrow -\infty$)
\begin{align}\label{48}
q_{1}&\rightarrow S_{1}^{1-}=\frac{a_{1}e^{\eta_{1}}}{1+e^{\eta^{*}_{1}+\eta_{1}+\theta_{1\bar{1}}}}
\rightarrow\frac{a_{1}}{2}e^{i\eta_{\scriptscriptstyle 1I}-\frac{\theta_{1\bar{1}}}{2}}\textup{sech}(\eta_{\scriptscriptstyle 1R}+\frac{\theta_{1\bar{1}}}{2}), \\
q_{2}&\rightarrow S_{2}^{1-}=\tau_{1}e^{i\psi}\frac{1+A_{1\bar{1}}e^{\eta_{1}+\eta^{*}_{1}}}{1+B_{1\bar{1}}e^{\eta_{1}+\eta^{*}_{1}}}
\rightarrow\frac{\tau_{1}}{2}e^{i\psi}[(1+e^{\phi_{1\bar{1}}})-(1-e^{\phi_{1\bar{1}}})\tanh(\eta_{\scriptscriptstyle 1R}+\frac{\theta_{1\bar{1}}}{2})],
\end{align}
\begin{figure}[!h]
\centering
\begin{minipage}{0.3\linewidth}\centering
\includegraphics[width=1\textwidth]{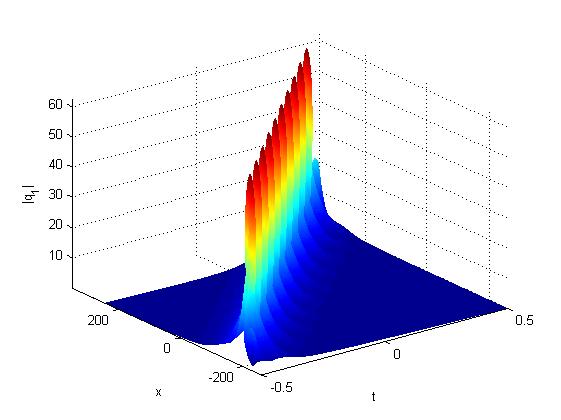}\caption*{(a1)}
\end{minipage}
\begin{minipage}{0.3\linewidth}\centering
\includegraphics[width=1\textwidth]{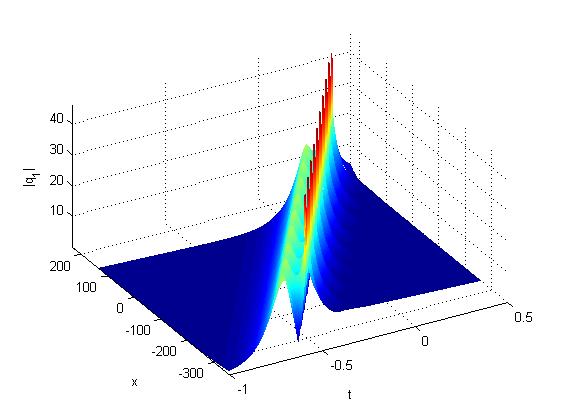}\caption*{(a2)}
\end{minipage}
\begin{minipage}{0.3\linewidth}\centering
\includegraphics[width=1\textwidth]{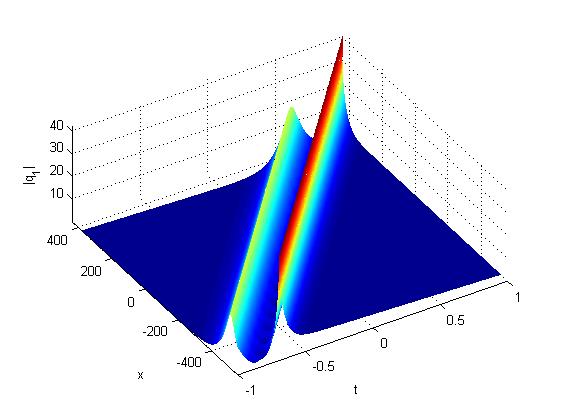}\caption*{(a3)}
\end{minipage}\\
\begin{minipage}{0.3\linewidth}\centering
\includegraphics[width=1\textwidth]{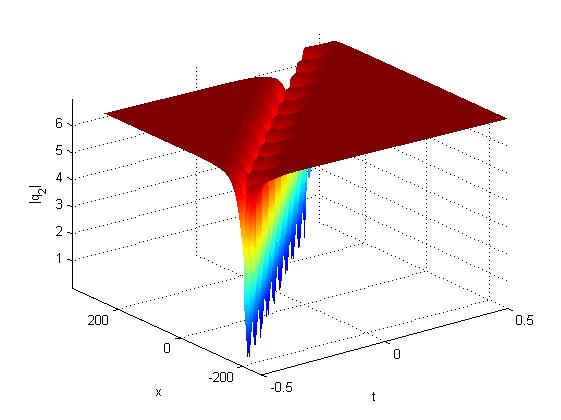}\caption*{(b1)}
\end{minipage}
\begin{minipage}{0.3\linewidth}\centering
\includegraphics[width=1\textwidth]{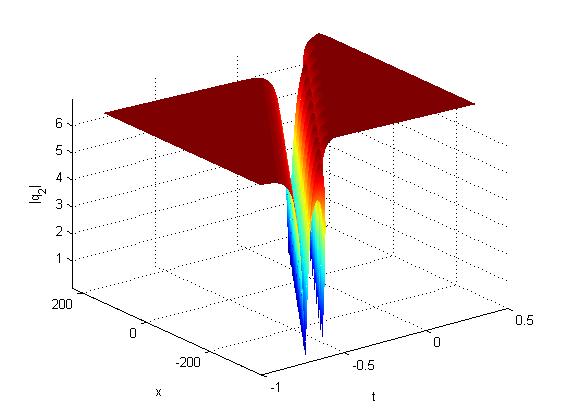}\caption*{(b2)}
\end{minipage}
\begin{minipage}{0.3\linewidth}\centering
\includegraphics[width=1\textwidth]{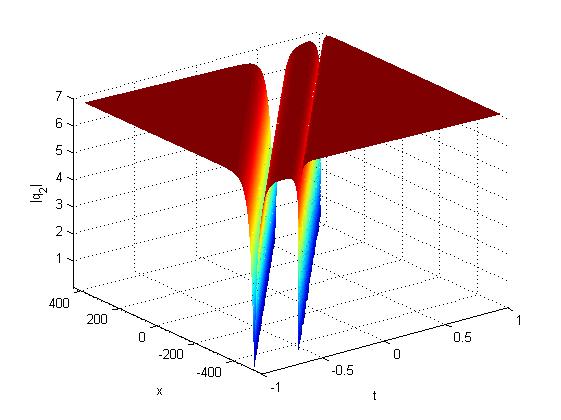}\caption*{(b3)}
\end{minipage}\\
\caption{\small{Bound states of solitons for solution\eqref{44}
(a1)-(b1) parameter $\tau_{1}=7$, $k=1, p_{1}=\frac{1}{2}+\frac{\sqrt{3}}{2}i, p_{2}=\frac{\sqrt{3}}{2}-\frac{1}{2}i, a_{1}=a_{2}=1, b_{1}=0, b_{2}=1$,
(a2)-(b2) with the same parameters as those in (a1-b1) except for $b_{1}=1, b_{2}=3$,
(a3)-(b3) with the same parameters as those in (a2-b2) except for $b_{1}=1, b_{2}=8$.
}}
\label{fig10}
\end{figure}

Soliton 2:(the wave-$\eta_{\rm\scriptscriptstyle 2R}$ is fixed, $\eta_{\scriptscriptstyle 1R}\rightarrow +\infty$)
\begin{align}\label{49}
q_{1}&\rightarrow S_{1}^{2-}=\frac{1}{2}e^{i\eta_{\scriptscriptstyle 2I}}e^{\theta_{_{1\bar{2}1}}-(\theta_{1\bar{1}2\bar{2}}+\theta_{1\bar{1}})/2}\textup{sech}(\eta_{\scriptscriptstyle 2R}+\frac{\theta_{1\bar{1}2\bar{2}}-\theta_{1\bar{1}}}{2}), \\
q_{2}&\rightarrow S_{2}^{2-}=\frac{\tau_{1}}{2}e^{i(\psi+\phi_{11})}[(1+e^{\phi_{2\bar{2}}})-(1-e^{\phi_{2\bar{2}}})\tanh(\eta_{\scriptscriptstyle 2R}+\frac{\theta_{1\bar{1}2\bar{2}}-\theta_{1\bar{1}}}{2})],
\end{align}
where
\begin{align}  \notag
G_{31}&=e^{\theta_{_{1\bar{2}1}}}=P_{12}(a_{1}P_{1\bar{1}}B_{2\bar{1}}-a_{2}P_{2\bar{1}}B_{1\bar{1}}),\\ \notag
\Omega&=e^{\theta_{1\bar{1}2\bar{2}}}=|P_{12}|^{2}(P_{1\bar{1}}P_{2\bar{2}}B_{1\bar{2}}
B_{2\bar{1}}-P_{1\bar{2}}P_{2\bar{1}}B_{1\bar{1}}B_{2\bar{2}}).
\end{align}
(ii) After collision $(t\rightarrow +\infty)$:\\
Soliton 1:(the wave-$\eta_{\rm\scriptscriptstyle 1R}$ is fixed, $\eta_{\scriptscriptstyle 2R}\rightarrow +\infty$)
\begin{align}\label{50}
q_{1}&\rightarrow S_{1}^{1+}=\frac{1}{2}e^{i\eta_{\scriptscriptstyle 1I}}e^{\theta_{_{1\bar{2}2}}-(\theta_{1\bar{1}2\bar{2}}+\theta_{2\bar{2}})/2}\textup{sech}(\eta_{\scriptscriptstyle 1R}+\frac{\theta_{1\bar{1}2\bar{2}}-\theta_{2\bar{2}}}{2}), \\
q_{2}&\rightarrow S_{2}^{1+}=\frac{\tau_{1}}{2}e^{i(\psi+\phi_{2\bar{2}})}[(1+e^{\phi_{1\bar{1}}})-
(1-e^{\phi_{1\bar{1}}})\tanh(\eta_{\scriptscriptstyle 1R}+\frac{\theta_{1\bar{1}2\bar{2}}-\theta_{2\bar{2}}}{2})],
\end{align}
where
\begin{align}  \notag
G_{32}=e^{\theta_{_{1\bar{2}2}}}=P_{12}(\alpha_{1}P_{1\bar{2}}B_{2\bar{2}}-\alpha_{2}P_{2\bar{2}}B_{1\bar{2}}).\\   \notag
\end{align}

Soliton 2:(the wave-$\eta_{\rm\scriptscriptstyle 2R}$ is fixed, $\eta_{\scriptscriptstyle 1R}\rightarrow -\infty$)
\begin{align}\label{51}
q_{1}&\rightarrow S_{1}^{2+}=\frac{a_{2}e^{\eta_{2}}}{1+e^{\eta_{2}+\eta_{2}+\theta_{2\bar{2}}}}
\rightarrow\frac{a_{2}}{2}e^{i\eta_{\scriptscriptstyle 2I}-\frac{\theta_{2\bar{2}}}{2}}\textup{sech}(\eta_{\scriptscriptstyle 2R}+\frac{\theta_{2\bar{2}}}{2}), \\
q_{2}&\rightarrow S_{2}^{2+}=\tau_{1}e^{i\psi}\frac{1+A_{2\bar{2}}e^{\eta_{2}+\eta^{*}_{2}}}{1+B_{2\bar{2}}e^{\eta_{2}+\eta^{*}_{2}}}
\rightarrow\frac{\tau_{1}}{2}e^{i\psi}[(1+e^{\phi_{2\bar{2}}})-(1-e^{\phi_{2\bar{2}}})\tanh(\eta_{\scriptscriptstyle 2R}+\frac{\theta_{2\bar{2}}}{2})].
\end{align}
with
\begin{align}
B_{i\bar{j}}=e^{\theta_{i\bar{j}}}~~(i=1,2) \notag
\end{align}

\begin{table}[htbp]
 \caption{Physical quantities of bright-dark soliton $S_{1}$ and $S_{2}$ before and after interactions.}
 \begin{tabular}{l}
  \toprule
  Soliton~~~~~~~~~Amplitude/depth~~~~~~~~~Velocity~~~~~~~~~~~Soliton~~~~~~~~~Amplitude/depth~~~~~~~~~~Velocity \\
   \midrule
  $S_{1}^{1-}$~~~~~~~~~~~~$\sqrt{\frac{a_{1}a_{1}^{*}}{4B_{1\bar{1}}}}$~~~~~~~~~~~~~~~~~~~~~~~~~$-\frac{|\tau_{1}|^{2}}{2|p_{1}|^{2}}$
  ~~~~~~~~~~~$S_{1}^{1+}$~~~~~~~~~~~~~~~~~~$\sqrt{\frac{G_{32}G_{32}^{*}}{4\Omega B_{2\bar{2}}}}$~~~~~~~~~~~~~~~~~$-\frac{|\tau_{1}|^{2}}{2|p_{1}|^{2}}$ \\

  $S_{2}^{1-}$~~~~~~~~~~~~~$|\tau_{1}|\sqrt{\frac{1+\mu_{1\bar{1}}}{2}\frac{\mu_{1\bar{1}}-1}{2}}$~~~~~~~~~~$-\frac{|\tau_{1}|^{2}}{2|p_{1}|^{2}}$
  ~~~~~~~~~~~~$S_{2}^{1+}$~~~~~~~~~~~~~~$|\tau_{1}|\sqrt{\frac{1+\mu_{1\bar{1}}}{2}\frac{\mu_{1\bar{1}}-1}{2}}$~~~~~~~~~~~$-\frac{|\tau_{1}|^{2}}{2|p_{1}|^{2}}$ \\

  $S_{1}^{2-}$~~~~~~~~~~~~~$\sqrt{\frac{G_{32}G_{32}^{*}}{4\Omega B_{1\bar{1}}}}$~~~~~~~~~~~~~~~~~~~~~$-\frac{|\tau_{1}|^{2}}{2|p_{2}|^{2}}$~~~~~~~~~~~~~$S_{1}^{2+}$~~~~~~~~~~~~~~~~~
  $\sqrt{\frac{a_{2}a_{2}^{*}}{4B_{2\bar{2}}}}$~~~~~~~~~~~~~~~~~~~~$-\frac{|\tau_{1}|^{2}}{2|p_{2}|^{2}}$ \\

  $S_{2}^{2-}$~~~~~~~~~~~~$|\tau_{1}|\sqrt{\frac{1+\mu_{2\bar{2}}}{2}\frac{\mu_{2\bar{2}}-1}{2}}$~~~~~~~~~~~~$-\frac{|\tau_{1}|^{2}}{2|p_{2}|^{2}}$
  ~~~~~~~~~~~$S_{2}^{2+}$~~~~~~~~~~~~~$|\tau_{1}|\sqrt{\frac{1+\mu_{2\bar{2}}}{2}\frac{\mu_{2\bar{2}}-1}{2}}$~~~~~~~~~~~~$-\frac{|\tau_{1}|^{2}}{2|p_{2}|^{2}}$ \\
  \bottomrule
 \end{tabular}
\end{table}

 Through the direct calculation, we get
\begin{equation*}
\frac{|a_{1}|}{\sqrt{B_{1\bar{1}}}}=\frac{|G_{32}|}{\sqrt{\Omega B_{2\bar{2}}}},~~
\frac{|a_{2}|}{\sqrt{B_{2\bar{2}}}}=\frac{|G_{31}|}{\sqrt{\Omega B_{1\bar{1}}}}.
\end{equation*}
From Table 2, we can see that the velocities and amplitude of both bright and dark solitons keep unchanged
before and after collisions except a phase shift.\\
\textbf{Remark 2}:~
For the one soliton solution, the parameter $p_{1}$ denotes the direction
of the soliton and $\tau_{1}$, $a_{1}$ give its amplitude. The one-soliton solution is
characterized by five parameters $\tau_{1},~a_{1},~b_{1},~k,~\lambda$. $\tau_{1}$
is restricted by the  parameters $k,~p_{1}$. Now the role of parameter $a_{1}$ can
be realized explicitly in the amplitude (intensity) of bright
component. This is shown in Fig.\eqref{fig7}.
The dark soliton part
influences the bright part through the parameters $\tau_{1}$(see Fig.\eqref{fig8}).
For the two bright-dark soliton, when $|p_{1}|=|p_{2}|$, the two solitons travel in parallel.
Due to the velocities of solitons and
distance between the solitons, oblique
interactions, attraction, exclusion soliton periodically will happen\cite{28}.
Fig. 9 and Fig. 10 show different kinds of soliton collisions.
\subsection{Dark-dark soliton solution}
In order to get dark-dark soliton solutions,
we assume $g_{j}=g_{0}^{(j)}(1+\chi g_{1}^{(j)}+\chi^{2} g_{2}^{(j)}+\cdots)~(j=1,2)$ and $f=1+
\chi f_{1}+\chi^{2} f_{2}+\cdots$ where $g_{k}^{(j)}~(k=1,2,\cdots)$
are complex functions and $f_{k}$ are real functions.
Substituting these forms into \eqref{39} and collecting the
coefficients of $\chi_{0}$, yields $g_{0}^{(j)}=\tau_{j}e^{i\psi_{j}}$,
$\psi_{j}=k_{j}y-\frac{\lambda}{k_{j}}s+\psi_{j}^{(0)}$ with $\lambda=\frac{1}{2}(|\tau_{2}|^{2}-|\tau_{1}|^{2})$,
in which $k_{j},\psi_{j}^{(0)}$ are real constants and $\tau_{j}$ are complex constants.\\
Eliminating $g_{0}^{(j)}$, we still use $g_{j}$ substituting $g_{j}$ for convenience.
Then the bilinear equation\eqref{39} can be rewritten as
\begin{align}\notag
&(D_{s}D_{y}-ik_{j}D_{s}+\frac{i\lambda}{k_{j}}D_{y})f\cdot g_{j}=0, \\  \label{54}
&(D_{s}^{2}-\lambda)f\cdot f=\frac{1}{2}(|\tau_{1}g_{1}|^{2}-|\tau_{2}g_{2}|^{2}),
\end{align}
where $g_{j}=1+\chi g_{1}^{(j)}+\chi^{2} g_{2}^{(j)}+\cdot\cdot\cdot$. Then the
equation \eqref{54} admits the solutions
\begin{equation}\label{55}
g_{1}^{(j)}=Z_{1}^{(j)}e^{\eta_{1}},~~f_{1}=e^{\eta_{1}},~~~\eta_{1}=P_{1}y-\Omega_{1}s+\eta^{(0)}_{1},
\end{equation}
in which $P_{1}$, $\Omega_{1}$, and $\eta^{(0)}_{1}$ are real constants and $Z_{1}^{(j)}$ are
complex constants. Collecting the
coefficients of $\chi_{1}$ from \eqref{54}, we have
\begin{equation}\label{56}
Z_{1}^{(j)}=\frac{-P_{1}\Omega_{1}+i(k_{j}\Omega_{1}+\frac{\lambda}{k_{j}}P_{1})}
{P_{1}\Omega_{1}+i(k_{j}\Omega_{1}+\frac{\lambda}{k_{j}}P_{1})}~(j=1,2),
\end{equation}
where
\begin{equation}\label{57}
\frac{|\tau_{2}|^{2}}{(P_{1}\Omega_{1})^{2}+(k_{2}\Omega_{1}+\frac{\lambda}{k_{2}}P_{1})^{2}}
-\frac{|\tau_{1}|^{2}}{(P_{1}\Omega_{1})^{2}+(k_{1}\Omega_{1}+\frac{\lambda}{k_{1}}P_{1})^{2}}=\frac{1}{P_{1}^{2}}.
\end{equation}
It's obvious that $|Z_{1}^{(j)}|=1$ and $g^{(j)}_{k}=f_{k}=0,(j=1,2,k=2,3,\cdots)$. Thus
the one dark soliton solution is written as
\begin{align}\label{58a}
q_{j}&=\frac{1}{2}\tau_{j}e^{i\psi_{j}}[(1+Z_{1}^{(j)})-(1-Z_{1}^{(j)})\tanh(\frac{\eta_{1}}{2})], \\  \label{58b}
x&=\lambda(y+s)+\frac{2\Omega_{1}e^{\eta_{1}}}{1+e^{\eta_{1}}},~~t=-s.
\end{align}
To analyze the dynamics of the one-soliton solution for the Eq.\eqref{9},
we should know the term $\partial x/\partial y$.
Although it is difficult to classify the types of soliton like the bright soliton
because the parameters $P_{1}$ and $\Omega_{1}$ can not be determined independently,
we can solve $\Omega_{1}, P_{1}$ in the special case $k_{j}=k(j=1,2)$.
Letting $Z_{1}^{(j)}=e^{2i\theta_{1}}$, it follows that
\begin{equation}\label{59}
\Omega_{1}=\sqrt{2\lambda}\sin{\theta_{1}},~~~P_{1}=
\frac{2k^{2}\sin{\theta_{1}}}{2k\cos{\theta_{1}}-\sqrt{2\lambda}}.
\end{equation}
Since
\begin{equation}\label{60}
\frac{\partial x}{\partial y}=\frac{\lambda+(2\lambda+  \frac{4\sqrt{2\lambda}k^{2}\sin^{2}{\theta_{1}}}
{2k\cos{\theta_{1}}-\sqrt{2\lambda}})e^{\eta_{1}}+\lambda e^{2\eta_{1}}}{(1+e^{\eta_{1}})^{2}},
\end{equation}
we classify this one-dark soliton solution as follows:\\
(i) when $\sqrt{2\lambda}-2k\cos{\theta_{1}}<0$, or $\sqrt{2\lambda}-2k\cos{\theta_{1}}>0$ and $\triangle_{1}>0$,
where $\triangle_{1}=\lambda-\sqrt{2\lambda}k\cos{\theta_{1}}-2k^{2}\sin^{2}{\theta_{1}} $
the single-dark soliton solution is always smooth.
An example is illustrated in Fig. 11(b1).\\
(ii) when $\sqrt{2\lambda}-2k\cos{\theta_{1}}>0$ and $\triangle_{1}=0$,
then the ${\partial x}/{\partial y}$ attains zero at only one point, which leads to a
cusponed dark soliton as displayed in Fig. 11(b2).\\
(iii) when $\sqrt{2\lambda}-2k\cos{\theta_{1}}>$ and $\triangle_{1}<0$,
then the ${\partial x}/{\partial y}$ attains zero at two point, which leads to a
loop dark soliton as displayed in Fig. 11(b3).\\
\begin{figure}[!h]
\centering
\begin{minipage}{0.3\linewidth}\centering
\includegraphics[width=1\textwidth]{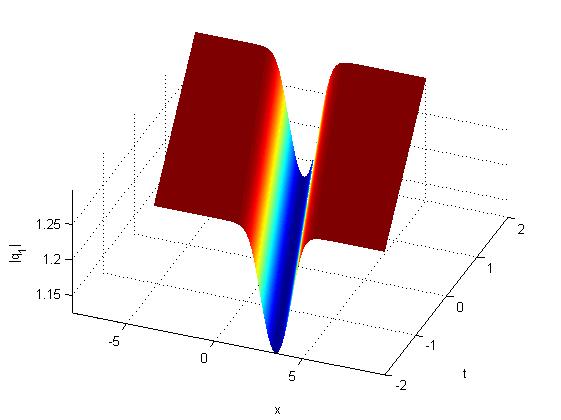}\caption*{(a1)}
\end{minipage}
\begin{minipage}{0.3\linewidth}\centering
\includegraphics[width=1\textwidth]{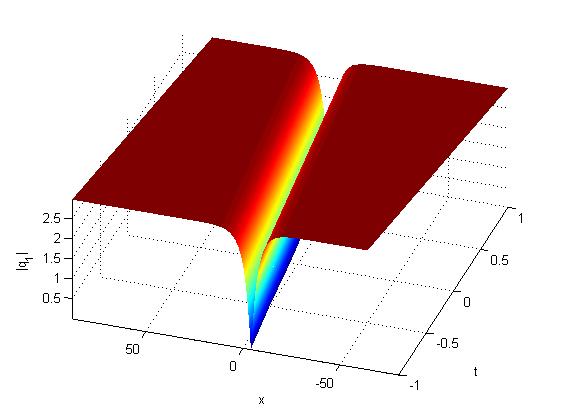}\caption*{(a2)}
\end{minipage}
\begin{minipage}{0.3\linewidth}\centering
\includegraphics[width=1\textwidth]{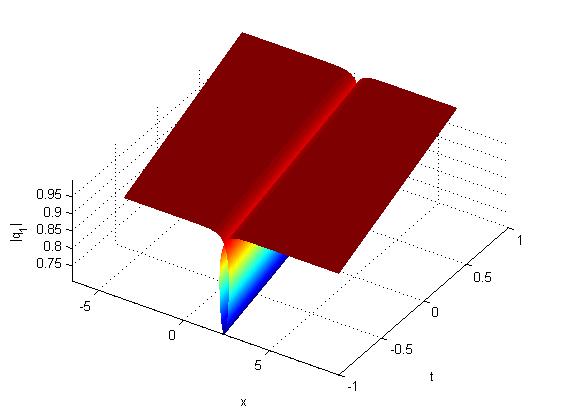}\caption*{(a3)}
\end{minipage}\\
\begin{minipage}{0.3\linewidth}\centering
\includegraphics[width=1\textwidth]{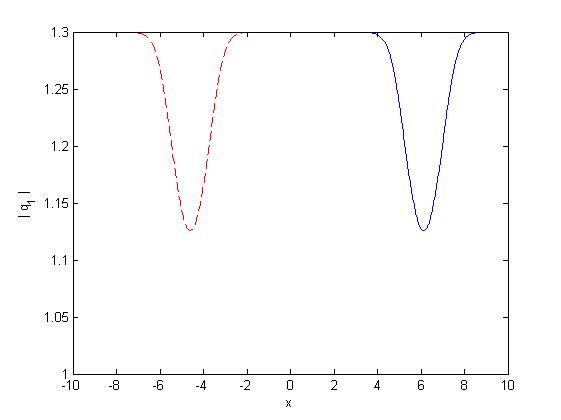}\caption*{(b1)}
\end{minipage}
\begin{minipage}{0.3\linewidth}\centering
\includegraphics[width=1\textwidth]{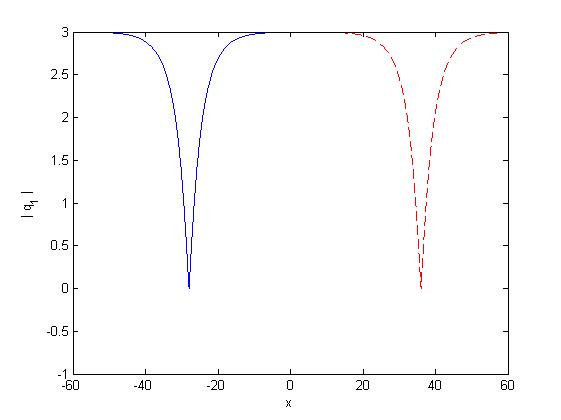}\caption*{(b2)}
\end{minipage}
\begin{minipage}{0.3\linewidth}\centering
\includegraphics[width=1\textwidth]{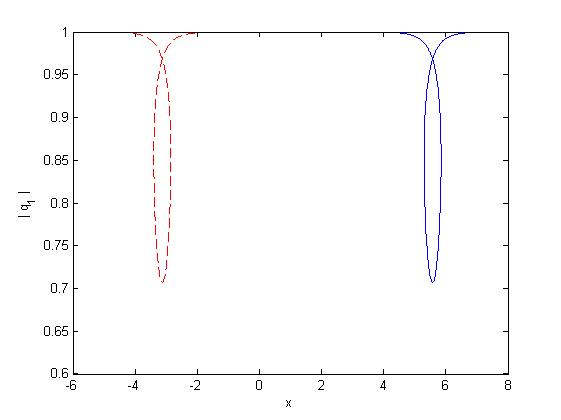}\caption*{(b3)}
\end{minipage}\\

\caption{\small{One dark soliton to the coupled focusing-defocusing complex short pulse equation: (a) 3D plot.
(b) profiles of $|q|$ at t = -4(blue solid line); t = 4(magenta dashed line).
(a1)-(b1) smooth soliton with $\tau_{1}=1.3$, $\tau_{2}=2$, $k=1,\theta_{1}=\frac{\pi}{6}$,~$\eta^{(0)}_{1}=0$,
(a2)-(b2) cuspon soliton with $\tau_{1}=3$, $\tau_{2}=5$, $k=2,\theta_{1}=\frac{\pi}{2}$,~$\eta^{(0)}_{1}=0$,
(a3)-(b3) loop soliton with $\tau_{1}=1$, $\tau_{2}=2$, $k=1,\theta_{1}=\frac{\pi}{4}$,~$\eta^{(0)}_{1}=0$.}}
\label{fig11}
\end{figure}

To construct the two dark-dark soliton solution, we assume
\begin{align}\label{61}
g_{1}^{(j)}&=\gamma_{1}^{(j)}e^{\eta_{1}}+\gamma_{2}^{(j)}e^{\eta_{2}},
f_{1}=e^{\eta_{1}}+e^{\eta_{2}}, \\ \notag
g_{2}^{(j)}&=\gamma_{1}^{(j)}\gamma_{2}^{(j)}\nu e^{\eta_{1}+\eta_{2}},
f_{2}=\nu e^{\eta_{1}+\eta_{2}}, \\ \notag
\eta_{j}&=P_{j}y-\Omega_{j}s+\eta^{(0)}_{j},~(j=1,2).\\  \notag
\end{align}
\begin{figure}[!h]
\centering
\begin{minipage}{0.3\linewidth}\centering
\includegraphics[width=1\textwidth]{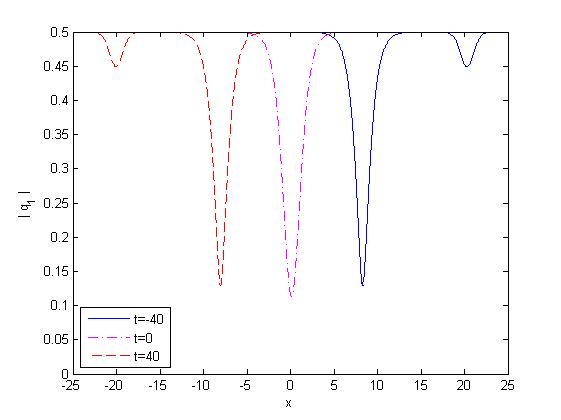}\caption*{(a1)}
\end{minipage}
\begin{minipage}{0.3\linewidth}\centering
\includegraphics[width=1\textwidth]{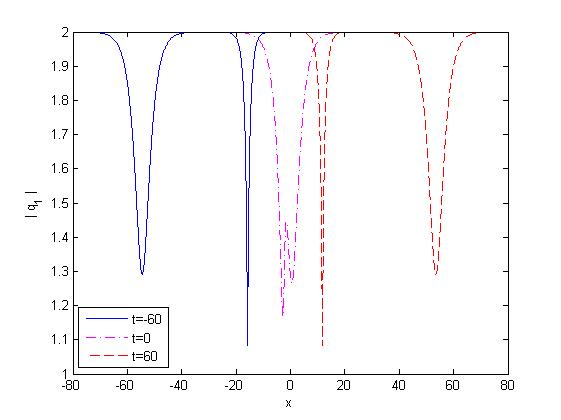}\caption*{(a2)}
\end{minipage}
\begin{minipage}{0.3\linewidth}\centering
\includegraphics[width=1\textwidth]{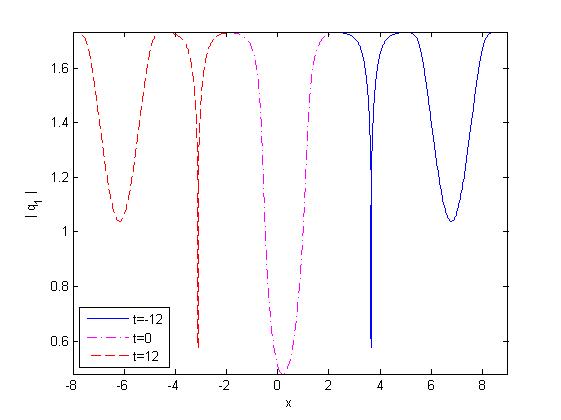}\caption*{(a3)}
\end{minipage}\\
\begin{minipage}{0.3\linewidth}\centering
\includegraphics[width=1\textwidth]{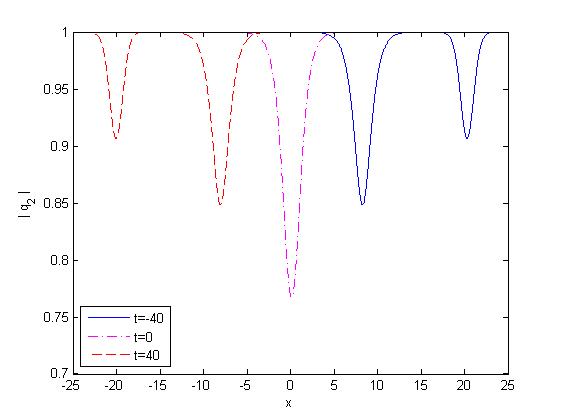}\caption*{(b1)}
\end{minipage}
\begin{minipage}{0.3\linewidth}\centering
\includegraphics[width=1\textwidth]{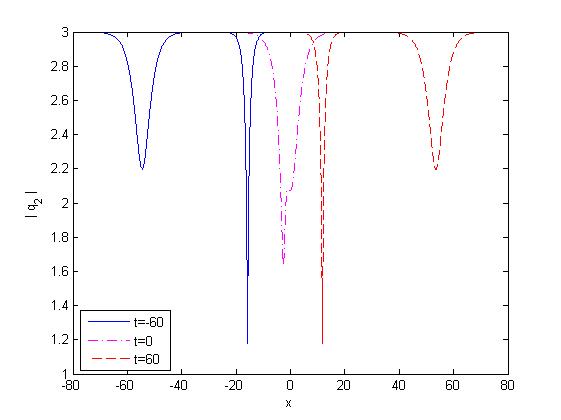}\caption*{(b2)}
\end{minipage}
\begin{minipage}{0.3\linewidth}\centering
\includegraphics[width=1\textwidth]{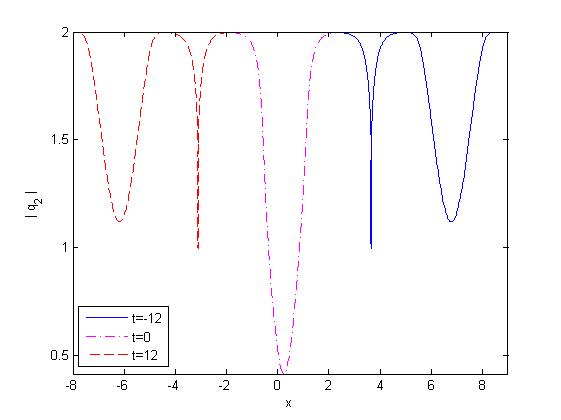}\caption*{(b3)}
\end{minipage}\\
\caption{\small{
Evolution plot of two dark-dark soliton solution for\eqref{61} with parameters $\eta^{(0)}_{1}=\eta^{(0)}_{2}=0$.
(a1)-(b1) parameters $\tau_{1}=\frac{1}{2}$, $\tau_{2}=1$, $k_{1}=1, k_{2}=\sqrt{2},P_{1}=1,
P_{2}=\frac{1}{2}, \Omega_{1}\approx0.35960053, \Omega_{2}\approx-0.21633656$,
(a2)-(b2) parameters $\tau_{1}=\sqrt{3}-i$, $\tau_{2}=\sqrt{5}+2i$, $k_{1}=1, k_{2}=2, P_{1}=2,
P_{2}=1, \Omega_{1}\approx-2.18581909, \Omega_{2}\approx-1.35443428$,
(a3)-(b3) parameters $\tau_{1}=1+\sqrt{2}i$, $\tau_{2}=1-\sqrt{3}i$, $k_{1}=1, k_{2}=1, P_{1}=2,
 P_{2}=5, \Omega_{1}\approx-0.58595167, \Omega_{2}\approx0.90979099$.}}
\label{fig12}
\end{figure}
Substituting \eqref{61} into \eqref{54} and collecting the terms with the same power of $\chi$, we have
\begin{align}\label{62}
\gamma_{l}^{(j)}&=\frac{-P_{l}\Omega_{l}+i(k_{j}\Omega_{l}+\frac{\lambda}{k_{j}}P_{l})}
{P_{l}\Omega_{l}+i(k_{j}\Omega_{l}+\frac{\lambda}{k_{j}}P_{l})},~(j,l=1,2)\\
\nu&=\frac{P_{1}P_{2}\Omega_{1}\Omega_{2}(P_{1}-P_{2})
(\Omega_{2}-\Omega_{1})+\lambda(P_{2}\Omega_{1}-P_{1}\Omega_{2})^{2}}
{-P_{1}P_{2}\Omega_{1}\Omega_{2}(P_{1}+P_{2})
(\Omega_{1}+\Omega_{2})+\lambda(P_{2}\Omega_{1}-P_{1}\Omega_{2})^{2}},
\end{align}
where $P_{j}$ and $\Omega_{j}$ satisfy
\begin{equation}\label{63}
\frac{|\tau_{2}|^{2}}{(P_{j}\Omega_{j})^{2}+(k_{2}\Omega_{j}+\frac{\lambda}{k_{2}}P_{j})^{2}}
-\frac{|\tau_{1}|^{2}}{(P_{j}\Omega_{j})^{2}+(k_{1}\Omega_{j}+\frac{\lambda}{k_{1}}P_{j})^{2}}=\frac{1}{P_{j}^{2}}~~(j=1,2).
\end{equation}
Then we get the dark-dark soliton solutions for Eq.\eqref{9}
\begin{align}\label{64a}
q_{j}&=\tau_{j}e^{i(k_{j}y-\frac{\lambda}{k_{j}}s)}\frac{1+\gamma_{1}^{(j)}e^{\eta_{1}}
+\gamma_{2}^{(j)}e^{\eta_{2}}+\gamma_{1}^{(j)}\gamma_{2}^{(j)}\nu e^{\eta_{1}+\eta_{2}}}
{1+e^{\eta_{1}}+e^{\eta_{2}}+\nu e^{\eta_{1}+\eta_{2}}},~~(j=1,2)\\                                                                           \label{64b}
x&=\lambda (y+s)+2\frac{\Omega_{1}e^{\eta_{1}}+\Omega_{2}e^{\eta_{2}}+(\Omega_{1}+\Omega_{2})\nu e^{\eta_{1}
+\eta_{2}}}{1+e^{\eta_{1}}+e^{\eta_{2}}+\nu e^{\eta_{1}+\eta_{2}}},~~~t=-s.
\end{align}
 To show that the solution indeed gives a dark-dark soliton for Eq.\eqref{9}, we consider its asymptotic behavior.
 We assume $P_{\rm\scriptscriptstyle 1R}, P_{\rm\scriptscriptstyle 2R},
 \Omega_{\rm\scriptscriptstyle 1R}, \Omega_{\rm\scriptscriptstyle 2R}>0$,
$\frac{P_{\rm\scriptscriptstyle 1R}}{P_{\rm\scriptscriptstyle 2R}}>
\frac{\Omega_{\rm\scriptscriptstyle 1R}}{\Omega_{\rm\scriptscriptstyle 2R}}$
without loss of generality. Then we discuss the following two cases: (i) when the wave-$\eta_{\rm\scriptscriptstyle 1R}$ is fixed,
$\eta_{\rm\scriptscriptstyle 2R}=\frac{P_{\rm\scriptscriptstyle 2R}}{P_{\rm\scriptscriptstyle 1R}}
\eta_{\rm\scriptscriptstyle 1R}-(\Omega_{\rm\scriptscriptstyle 2R}-
\frac{P_{\rm\scriptscriptstyle 2R}}{P_{\rm\scriptscriptstyle 1R}}\Omega_{\rm\scriptscriptstyle 1R})s$.
When $t\rightarrow \pm\infty$, we have $\eta_{\rm\scriptscriptstyle 2R}\rightarrow \pm\infty$  for soliton 1.
(ii) When the wave-$\eta_{\rm\scriptscriptstyle 2R}$ is fixed,
$\eta_{\rm\scriptscriptstyle 1R}=\frac{P_{\rm\scriptscriptstyle 1R}}{P_{\rm\scriptscriptstyle 2R}}
\eta_{\rm\scriptscriptstyle 2R}-(\Omega_{\rm\scriptscriptstyle 1R}-
\frac{P_{\rm\scriptscriptstyle 1R}}{P_{\rm\scriptscriptstyle 2R}}\Omega_{\scriptscriptstyle 2R})s$.
When $t\rightarrow \pm\infty$, $\eta_{\rm\scriptscriptstyle 1R}\rightarrow \mp\infty$ for soliton 2.
This leads to the following asymptotic forms for two-soliton solution under the reciprocal transformation\eqref{64b}.\\
(i)Soliton 1:(the wave-$\eta_{\rm\scriptscriptstyle 1R}$ is fixed, $\eta_{\rm\scriptscriptstyle 2R}\rightarrow \mp\infty$),
\begin{equation}\label{65a}
q_{j}\sim \left\{
\begin{aligned}
&\frac{1}{2}\tau_{j}e^{i\psi_{j}}[(1+\gamma_{1}^{(j)})-
(1-\gamma_{1}^{(j)})\tanh(\frac{\eta_{1}}{2})] ~~~(t\rightarrow -\infty),\\
&\frac{1}{2}\tau_{j}\gamma_{2}^{(j)}e^{i\psi_{j}} [(1+\gamma_{1}^{(j)})
-(1-\gamma_{1}^{(j)})\tanh(\frac{\eta_{1}}{2}+\eta_{10})]~~~(t\rightarrow +\infty),
\end{aligned}
\right.
\end{equation}
(ii)Soliton 2:(the wave-$\eta_{\rm\scriptscriptstyle 2R}$ is fixed,, $\eta_{\rm\scriptscriptstyle 1R}\rightarrow \pm\infty$),
\begin{equation}\label{65b}
q_{j}\sim \left\{
\begin{aligned}
&\frac{1}{2}\tau_{j}\gamma_{1}^{(j)}e^{i\psi_{j}}[(1+\gamma_{2}^{(j)})
-(1-\gamma_{2}^{(j)})\tanh(\frac{\eta_{2}}{2}+\eta_{10})]~~~(t\rightarrow -\infty),\\
&\frac{1}{2}\tau_{j}e^{i\psi_{j}}[(1+\gamma_{2}^{(j)})-
(1-\gamma_{2}^{(j)})\tanh(\frac{\eta_{2}}{2})] ~~~(t\rightarrow +\infty).
\end{aligned}
\right.
\end{equation}
\textbf{Remark 3}:
From the analysis of the asymptotic behavior of
the two dark-dark solitons,  we can see that the interaction of the
two solitons are elastic.
The collision processes between smooth-smooth dark solitons,
smooth-cuspon dark solitons and smooth-loop dark solitons are
illustrated in Fig.11(a1-a3), respectively.
When a smooth dark soliton interacts with a
cuspon dark soliton, the singularity of the cuspon dark
soliton still maintains as observed in Fig.12(a2-b2) and Fig.12(a3-b3),
which is different from the interaction of the dark soliton of defocusing CSP equation\cite{22}.\\
Especially, when $k_{j}=k (j=1,2,\cdots N)$, the explicit
N-dark soliton solutions for Eq.\eqref{9} can be written as
\begin{align}\label{65}
q_{1,N}&=\tau_{1}e^{i(ky-\frac{\lambda}{k}s)} \frac{\sum_{\mu=0,1}e^{\sum^{n}_{j=1}\mu_{j}
(\eta_{j}+2i\theta_{j})}+\sum^{n}_{1\leq j<l}{\mu_{j}\mu_{l}A_{jl}}}
{\sum_{\mu=0,1}e^{\sum^{n}_{j=1}\mu_{j}\eta_{j}}+\sum^{n}_{1\leq j<l}{\mu_{j}\mu_{l}A_{jl}}},\\    \notag
q_{2,N}&=\tau_{2}e^{i(ky-\frac{\lambda}{k}s)} \frac{\sum_{\mu=0,1}e^{\sum^{n}_{j=1}\mu_{j}
(\eta_{j}+2i\theta_{j})}+\sum^{n}_{1\leq j<l}{\mu_{j}\mu_{l}A_{jl}}}
{\sum_{\mu=0,1}e^{\sum^{n}_{j=1}\mu_{j}\eta_{j}}+\sum^{n}_{1\leq j<l}{\mu_{j}\mu_{l}A_{jl}}},
\end{align}
where
\begin{align}\notag
\eta_{j}&=P_{j}y-\Omega_{j}s+\eta^{(0)}_{j},e^{A_{jl}}=
\left(\frac{\sin{\frac{\theta_{j}-\theta_{l}}{l}}}{\sin{\frac{\theta_{j}+\theta_{l}}{l}}}\right)^{2},\\ \label{66}
\Omega_{j}&=\sqrt{|\tau_{2}|^{2}-|\tau_{1}|^{2}}\sin{\theta_{j}},~~~P_{j}
=\frac{2k^{2}\sin{\theta_{j}}}{2k\cos{\theta_{j}}-\sqrt{|\tau_{2}|^{2}-|\tau_{1}|^{2}}},\\   \notag
x&=\lambda(y+s)-2[\ln({\sum_{\mu=0,1}e^{\sum^{n}_{j=1}\mu_{j}\eta_{j}}+\sum^{n}_{1\leq j<l}{\mu_{j}\mu_{l}A_{jl}}})]_{s},~~~t=-s.
\end{align}

\section{Breather and rogue wave solutions}
\subsection{Breather solution}

\begin{figure}[!h]
\centering
\begin{minipage}{0.3\linewidth}\centering
\includegraphics[width=1\textwidth]{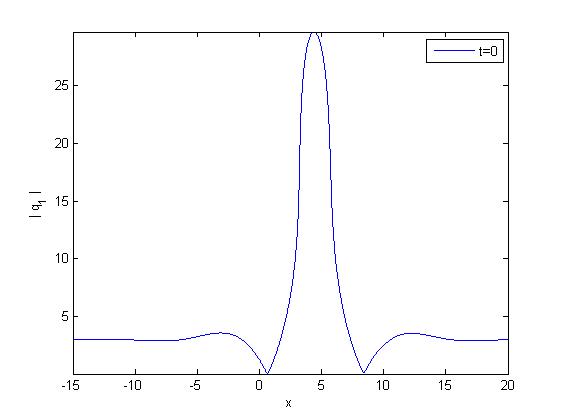}\caption*{(a1)}
\end{minipage}
\begin{minipage}{0.3\linewidth}\centering
\includegraphics[width=1\textwidth]{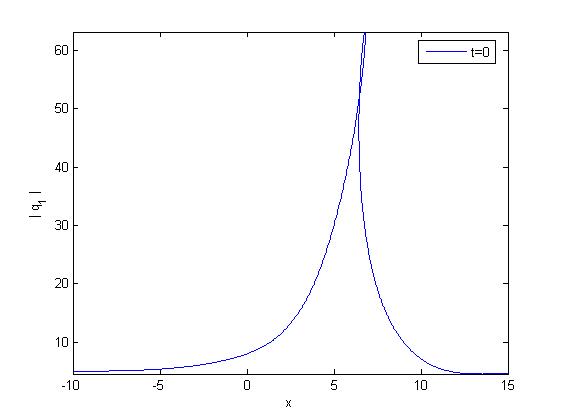}\caption*{(a2)}
\end{minipage}
\begin{minipage}{0.3\linewidth}\centering
\includegraphics[width=1\textwidth]{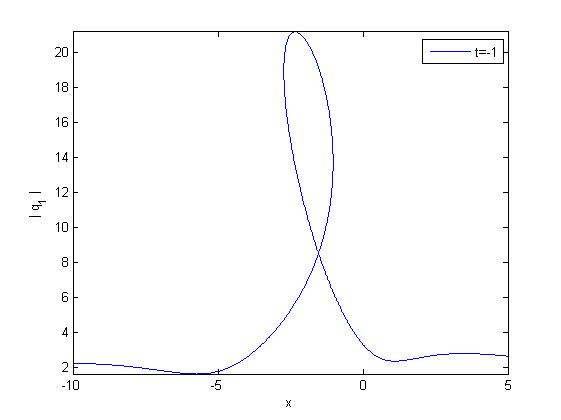}\caption*{(a3)}
\end{minipage}\\
\begin{minipage}{0.3\linewidth}\centering
\includegraphics[width=1\textwidth]{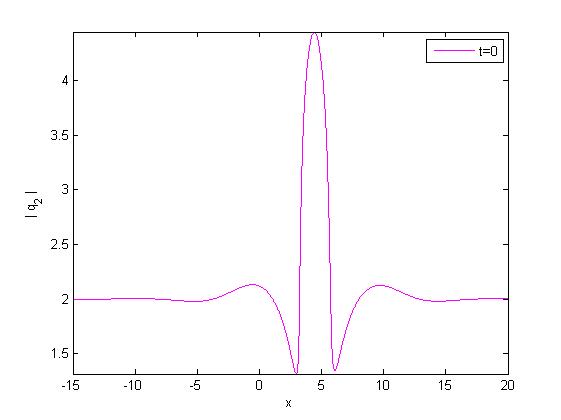}\caption*{(b1)}
\end{minipage}
\begin{minipage}{0.3\linewidth}\centering
\includegraphics[width=1\textwidth]{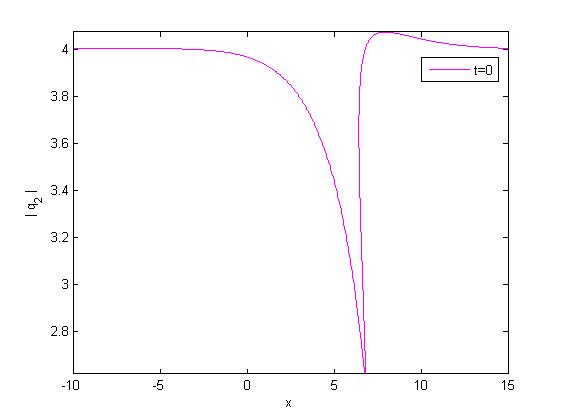}\caption*{(b2)}
\end{minipage}
\begin{minipage}{0.3\linewidth}\centering
\includegraphics[width=1\textwidth]{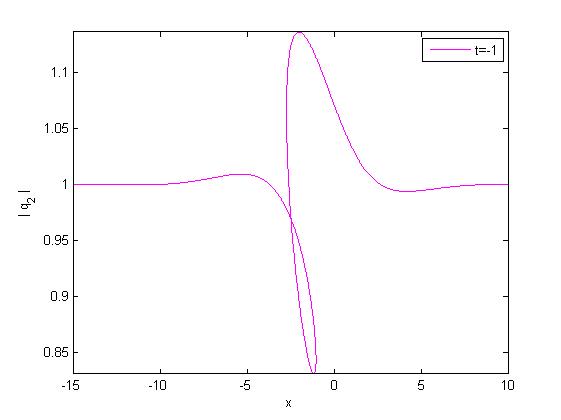}\caption*{(b3)}
\end{minipage}\\
\caption{\small{Evolution breathers for Eq.\eqref{9}.
(a1)-(b1)Smooth breather with parameters $\tau_{1}=3, \tau_{2}=2, k_{1}=1, k_{2}=2, P_{1}=1+\sqrt{3}i,
\Omega_{1}\approx 1.54849728-5.02607499i, \eta^{(0)}=0$,
(a2)-(b2)Breather withparameters $\tau_{1}=5, \tau_{2}=4, k_{1}=\frac{1}{2}, k_{2}=\frac{1}{4},
P_{1}=2-i, \Omega_{1}\approx 1.43479848-9.30893675i, \eta^{(0)}=0$,
(a3)-(b3)loop breather with parameters $\tau_{1}=1+2i, \tau_{2}=1, k_{1}=\frac{1}{2}, k_{2}=\frac{1}{4},
P_{1}=1+i, \Omega_{1}\approx 1.60552879-5.39497000i, \eta^{(0)}=0$.}}
\label{fig13}
\end{figure}
We know the breather soliton solution can be derived from
dark-dark soliton solution. In this section, we will give
the breather soliton to Eq.\eqref{9} from Eq.\eqref{64a}.
We assume $\Omega_{j}, \eta^{(0)}_{j}$
are complex and $k_{j}~(j=1,2)$ are still real constants,
and set $\eta_{2}=\eta_{1}^{*}=\eta$ and $\phi_{2}=\phi_{1}^{*}\pm\pi$, $\varphi_{2}=\varphi_{1}^{*}\pm\pi$.
$\phi_{1}=\phi_{\scriptscriptstyle R}+i\phi_{\scriptscriptstyle I}$,
$\varphi_{1}=\varphi_{\scriptscriptstyle R}+i\varphi_{\scriptscriptstyle I}$.
Then $q_{1}, q_{2}$ can be rewritten as
\begin{align}\notag
q_{1}&=\tau_{1}e^{i(\psi_{1}+2\phi_{\scriptscriptstyle R})}\frac{\sqrt{\nu}\cosh(\eta_{\scriptscriptstyle R}+2i\phi_{\scriptscriptstyle R}+\sigma)+\cos({\eta_{\scriptscriptstyle I}+2i\phi_{\scriptscriptstyle I}})}
{\sqrt{\nu}\cosh(\eta_{\scriptscriptstyle R}+\sigma)+\cos{\eta_{\scriptscriptstyle I}}},\\ \label{67}
q_{2}&=\tau_{2}e^{i(\psi_{2}+2\varphi_{\scriptscriptstyle R})}
\frac{\sqrt{\nu}\cosh(\eta_{\scriptscriptstyle R}+2i\varphi_{\scriptscriptstyle R}+\sigma)+\cos{(\eta_{\scriptscriptstyle I}+2i\varphi_{\scriptscriptstyle I})}}{\sqrt{\nu}\cosh(\eta_{\scriptscriptstyle R}+\sigma)+\cos{\eta_{\scriptscriptstyle I}}},\\   \notag
x&=\lambda (y+s)+\frac{2\sqrt{\nu}\sinh(\eta_{\scriptscriptstyle R}+\sigma)-\Omega_{\scriptscriptstyle I}\sin{\eta_{\scriptscriptstyle I}}}{\sqrt{\nu}\cosh(\eta_{\scriptscriptstyle R}+\sigma)+\cos{\eta_{\scriptscriptstyle I}}},~~t=-s.
\end{align}
where $\eta=\eta_{\scriptscriptstyle R}+i\eta_{\scriptscriptstyle I}$, $P_1=P_{\scriptscriptstyle R}+iP_{\scriptscriptstyle I}, \Omega_1=\Omega_{\scriptscriptstyle R}+i\Omega_{\scriptscriptstyle I}$,
$\eta_{\scriptscriptstyle R}=P_{\scriptscriptstyle R}y-\Omega_{\scriptscriptstyle R}s+\eta_{\scriptscriptstyle R}^{(0)}, \eta_{I}=P_{\scriptscriptstyle I}y-\Omega_{\scriptscriptstyle I}s+\eta_{\scriptscriptstyle I}^{(0)}$,
$\sigma=\frac{1}{2}\ln{\nu}$, $\phi_{\scriptscriptstyle R}=\ln|\gamma_{1}^{(1)}|, \phi_{\scriptscriptstyle I}=\arg(\gamma_{1}^{(1)})$, $\varphi_{\scriptscriptstyle R}=\ln|\gamma_{1}^{(2)}|, \varphi_{\scriptscriptstyle I}=\arg(\gamma_{1}^{(2)})$, where $P_{1},\Omega_{1}$ and $\nu$ satisfy
\begin{align}\notag
\lambda&=\frac{|\tau_{2}|^{2}-|\tau_{1}|^{2}}{2},~~~\gamma_{1}^{(j)}=
\frac{-P_{1}\Omega_{1}+i(k_{j}\Omega_{1}+\frac{\lambda}{k_{j}}P_{1})}
{P_{1}\Omega_{1}+i(k_{j}\Omega_{1}+\frac{\lambda}{k_{j}}P_{1})},~~(j=1,2),\\         \label{68}
\nu&=\frac{-P_{I}\Omega_{I}|P_{1}\Omega_{1}|^{2}
+\lambda(P_{R}\Omega_{I}-P_{I}\Omega_{R})^{2}}
{P_{R}\Omega_{R}|P_{1}\Omega_{1}|^{2}+\lambda(P_{R}\Omega_{I}-P_{I}\Omega_{R})^{2}},\\  \notag
\end{align}
 \begin{figure}[!h]
\centering
\begin{minipage}{0.3\linewidth}\centering
\includegraphics[width=1\textwidth]{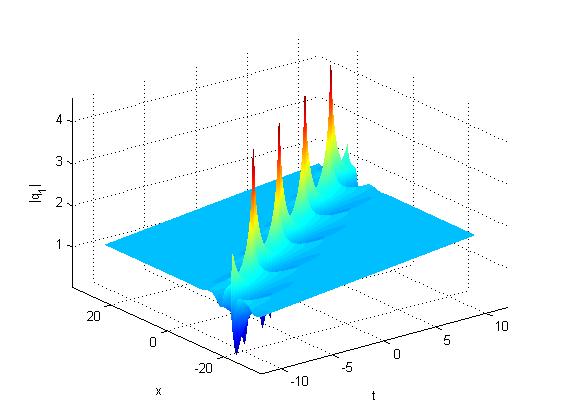}\caption*{(a1)}
\end{minipage}
\begin{minipage}{0.3\linewidth}\centering
\includegraphics[width=1\textwidth]{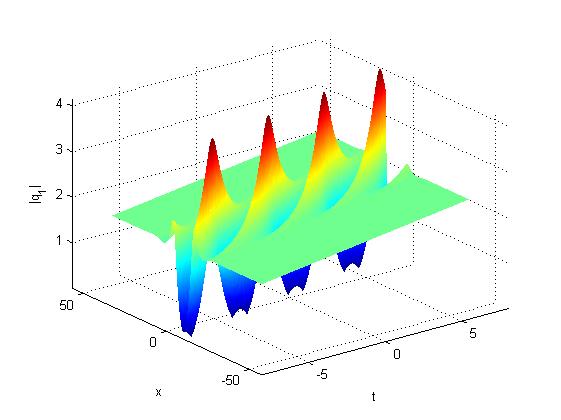}\caption*{(a2)}
\end{minipage}
\begin{minipage}{0.3\linewidth}\centering
\includegraphics[width=1\textwidth]{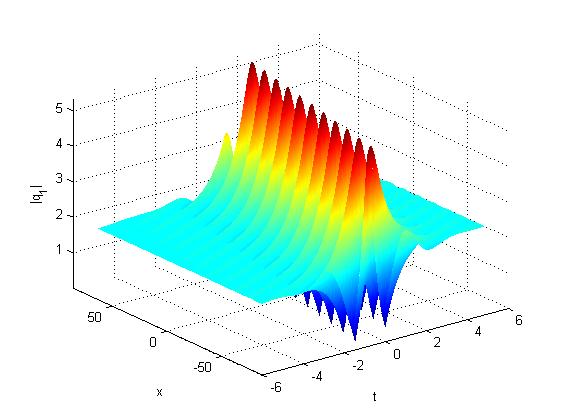}\caption*{(a3)}
\end{minipage}\\
\begin{minipage}{0.3\linewidth}\centering
\includegraphics[width=1\textwidth]{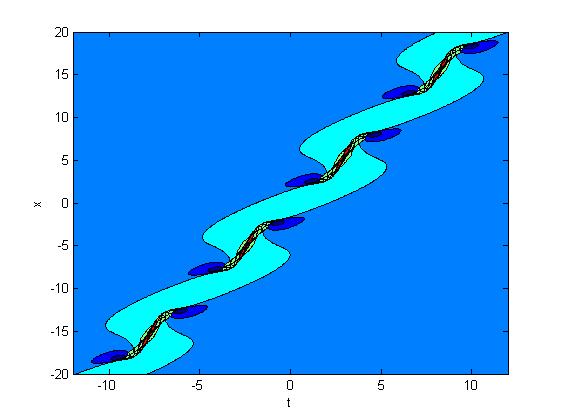}\caption*{(b1)}
\end{minipage}
\begin{minipage}{0.3\linewidth}\centering
\includegraphics[width=1\textwidth]{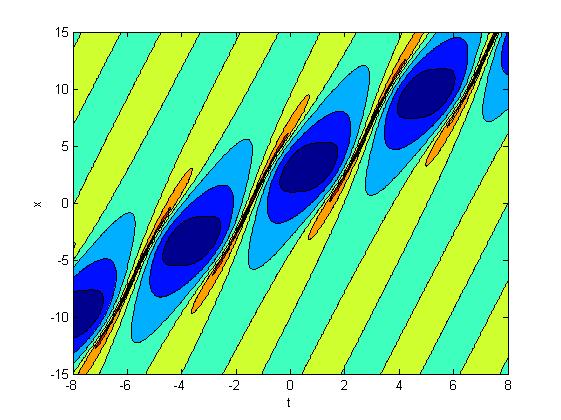}\caption*{(b2)}
\end{minipage}
\begin{minipage}{0.3\linewidth}\centering
\includegraphics[width=1\textwidth]{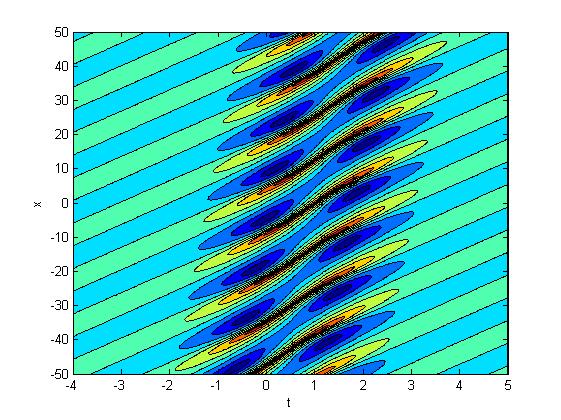}\caption*{(b3)}
\end{minipage}\\
\caption{\small{3D and contour plots of the breather soliton for the Eq.\eqref{9} with parameter $k=1, \eta^{(0)}=0$.
(a1) parameter with $\tau_{1}=1.4, \tau_{2}=1, \theta_{R}=0, \theta_{I}=\frac{1}{2}$,
(a2) parameter with $\tau_{1}=2, \tau_{2}=1, \theta_{R}=1, \theta_{I}=0$,
(a3) parameter with $\tau_{1}=2, \tau_{2}=1, \theta_{R}=1, \theta_{I}=1$}}
\label{fig14}
\end{figure}
\begin{equation}\label{69}
\frac{1}{P_{1}^{2}}=\frac{|\tau_{2}|^{2}}{(P_{1}\Omega_{1})^{2}+(k_{2}\Omega_{1}+\frac{\lambda}{k_{2}}P_{1})^{2}}
-\frac{|\tau_{1}|^{2}}{(P_{1}\Omega_{1})^{2}+(k_{1}\Omega_{1}+\frac{\lambda}{k_{1}}P_{1})^{2}}.
\end{equation}
Especially, when $k_{j}=k (j=1,2)$, we rewrite $q_1, q_2$ as
\begin{align}  \notag
q_{1}&=\tau_{1}e^{i(\psi+2\theta_{\rm\scriptscriptstyle R})}
\frac{\sqrt{\nu}\cosh(\eta_{\rm\scriptscriptstyle R}+
2i\theta_{\rm\scriptscriptstyle R}+\sigma)+
\cos({\eta_{\rm\scriptscriptstyle I}+2i\theta_{\rm\scriptscriptstyle I}})}
{\sqrt{\nu}\cosh(\eta_{\rm\scriptscriptstyle R}+\sigma)+
\cos{\eta_{\rm\scriptscriptstyle I}}},\\   \label{70a}
q_{2}&=\tau_{2}e^{i(\psi+2\theta_{\rm\scriptscriptstyle R})}
\frac{\sqrt{\nu}\cosh(\eta_{\rm\scriptscriptstyle R}+
2i\theta_{\rm\scriptscriptstyle R}+\sigma)+\cos{(\eta_{\rm\scriptscriptstyle I}
+2i\theta_{\rm\scriptscriptstyle I})}}
{\sqrt{\nu}\cosh(\eta_{\rm\scriptscriptstyle R}+\sigma)
+\cos{\eta_{\rm\scriptscriptstyle I}}}, \\ \label{70b}
x&=\lambda (y+s)+\frac{2\sqrt{\nu}\sinh(\eta_{\rm\scriptscriptstyle R}+\sigma)
-\Omega_{\rm\scriptscriptstyle I}\sin{\eta_{\rm\scriptscriptstyle I}}}
{\sqrt{\nu}\cosh(\eta_{\rm\scriptscriptstyle R}+\sigma)+
\cos{\eta_{\rm\scriptscriptstyle I}}},~~t=-s,
\end{align}
where
\begin{align}\notag
\Delta&=\sqrt{|\tau_{1}|^{2}-|\tau_{2}|^{2}},~~~\nu=\frac{\cosh^{2}{\theta_{\scriptscriptstyle I}}}{\cos^{2}{\theta_{\scriptscriptstyle R}}},~\\ \notag
 \Omega_{\rm\scriptscriptstyle R}&=-\Delta\cos{\theta_{\rm\scriptscriptstyle R}}\sinh{\theta_{\rm\scriptscriptstyle I}},~~~
 \Omega_{\rm\scriptscriptstyle I}=\Delta\sin{\theta_{\rm\scriptscriptstyle R}}\cosh{\theta_{\rm\scriptscriptstyle I}},\\ \label{71}
 P_{\scriptscriptstyle R}&=\frac{2k^{2}\cos{\theta_{\scriptscriptstyle R}}(2k\sin{\theta_{\scriptscriptstyle R}}-\Delta\sinh{\theta_{\rm\scriptscriptstyle I}}     )}{(\Delta+2k\sin{\theta_{\rm\scriptscriptstyle R}}\sinh{\theta_{\rm\scriptscriptstyle I}})^2+(2k\cos{\theta_{\rm\scriptscriptstyle R}}\cosh{\theta_{\rm\scriptscriptstyle I}})^2},\\  \notag
  P_{\rm\scriptscriptstyle I}&=\frac{2k^{2}\cosh{\theta_{\rm\scriptscriptstyle I}}(\Delta\sin{\theta_{\rm\scriptscriptstyle R}}+2k\sinh{\theta_{\rm\scriptscriptstyle I}})}{(\Delta+2k\sin{\theta_{\rm\scriptscriptstyle R}}\sinh{\theta_{\rm\scriptscriptstyle I}})^2+(2k\cos{\theta_{\rm\scriptscriptstyle R}}\cosh{\theta_{\rm\scriptscriptstyle I}})^2}.
\end{align}
\textbf{Remark 4}: Fig.13 illustrates the 2D plots of the
breather soliton solution\eqref{67} at time $t=0$.
From the Figs.14(a) and 14(b), we observe
that due to the hodograph transformation\eqref{70b},
when $\theta_{\rm\scriptscriptstyle 1R}=0$ or $\theta_{\rm\scriptscriptstyle 1I}=0$, the Akhmediev breather and Ma
breather are no longer existence which is different from the GCNLS\cite{33}.
\subsection{Rogue wave soliton solutions}
\begin{figure}[!h]
\centering
\begin{minipage}{0.3\linewidth}\centering
\includegraphics[width=1\textwidth]{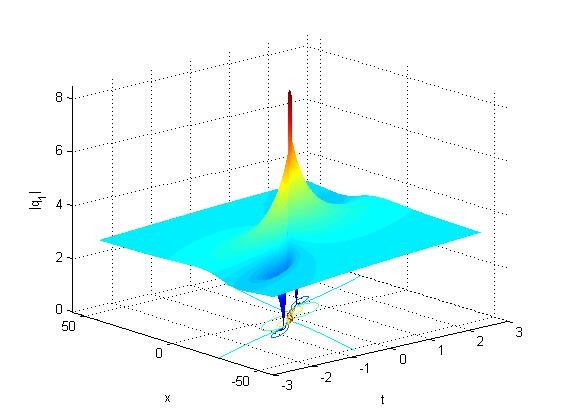}\caption*{(a1)}
\end{minipage}
\begin{minipage}{0.3\linewidth}\centering
\includegraphics[width=1\textwidth]{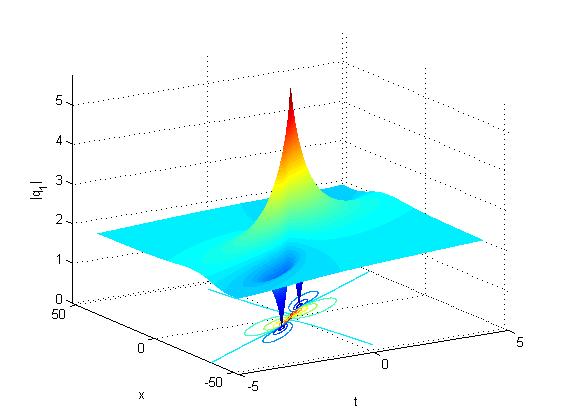}\caption*{(a2)}
\end{minipage}\\
\begin{minipage}{0.3\linewidth}\centering
\includegraphics[width=1\textwidth]{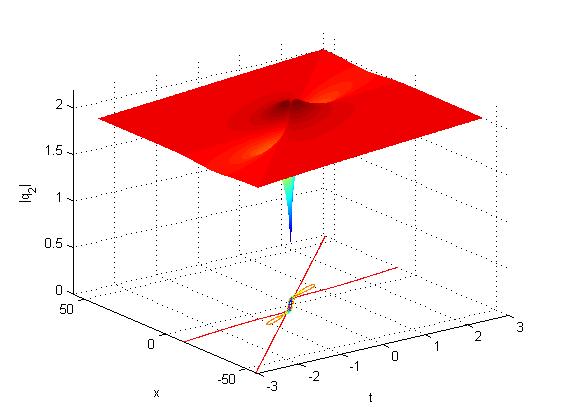}\caption*{(b1)}
\end{minipage}
\begin{minipage}{0.3\linewidth}\centering
\includegraphics[width=1\textwidth]{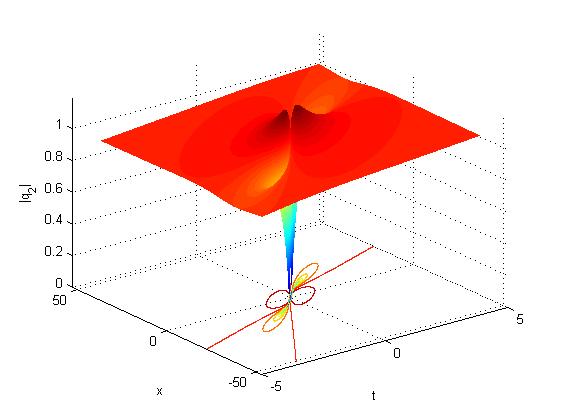}\caption*{(b2)}
\end{minipage}\\
\caption{\small{Rogue wave for Eq.\eqref{9} with the contour line drawn at the bottom of the figure.
(a1)-(b1)~$q_{1}$ is a bright rogue wave.~$q_{2}$ is a dark rogue wave with parameter
$\tau_{1}=3, \tau_{2}=2, k_{1}=1, k_{2}=2, \omega_{1}\thickapprox2.60485+2.95036i$.
(a2)-(b2)~$q_{1}$ and $q_{2}$ are four-petals rogue wave with parameter
$\tau_{1}=2, \tau_{2}=1, k_{1}=\frac{2}{3}, k_{2}=1, \omega_{1}\thickapprox3.479849-2.95036i$. }}
\label{fig15}
\end{figure}
The rogue wave appeared in the nonlinear system has been widely studied\cite{35,36,37,38,39,40}.
In order to construct the rogue wave solution to coupled focusing-defocusing complex short pulse equation\eqref{9}, we consider the breather solution \eqref{67}.
Let $P_1=\varepsilon$ and $\varepsilon$ is a complex small parameter. From the expression\eqref{69},
we have $\Omega_1=\omega_{1}\varepsilon+o(\varepsilon)$, where $\omega_{1}$ satisfies
\begin{equation}\label{72}
|k_2\tau_2|^{2}(\lambda+\omega_{1}k_1^{2})^2-(\lambda+\omega_{1}k_2^{2})^2(|k_1\tau_1|^{2}+(\lambda+\omega_{1}k_1^{2})^2)=0
\end{equation}
If $\eta_1^{(0)}=i\pi$, then $\eta_1=\varepsilon(x-\omega_{1}t)+i\pi+o(\varepsilon^2)$, and
\begin{align*}
\gamma_{1}^{(j)}&=1+\frac{2ik_j\omega_{1}\varepsilon}{\lambda+k_j^{2}\omega_{1}}
-\frac{2(k_j\omega_{1})^{2}\varepsilon^{2}}{(\lambda+k_j^{2}\omega_{1})^{2}}+O(\varepsilon^{2}),\\
\gamma_{2}^{(j)}&=1+\frac{2ik_j\omega_{1}^{*}\varepsilon^{*}}{\lambda+k_j^{2}\omega_{1}^{*}}
-\frac{2(k_j\omega_{1}^{*})^{2}\varepsilon^{*2}}{(\lambda+k_j^{2}\omega_{1}^{*})^{2}}+O(\varepsilon^{*2}),\\
\nu&=1+\frac{2(\omega_{1}+\omega_{1}^{*})|\omega_{1}|^2|\varepsilon|^2}{\lambda(\omega_{1}-\omega_{1}^{*})^{2}}
+O(\varepsilon^{2},\varepsilon^{*2},|\varepsilon|^{2}).
\end{align*}
We thus have,
\begin{align}\label{73}  \notag
f=&(|y-\omega_{1}s|^2+\frac{2(\omega_{1}+\omega_{1}^{*})|\omega_{1}|^2}{\lambda(\omega_{1}-\omega_{1}^{*})^{2}})
|\varepsilon|^{2}+O(\varepsilon^{2},\varepsilon^{*2},|\varepsilon|^{2}),\\
g_{j}=&(|y-\omega_{1}s|^2+\frac{2(\omega_{1}+\omega_{1}^{*})|\omega_{1}|^2}{\lambda(\omega_{1}-\omega_{1}^{*})^{2}}
+\frac{2ik_j\omega_{1}(y-\omega_{1}^{*}s)}{\lambda+k_j^{2}\omega_{1}}+
\frac{2ik_j\omega_{1}^{*}(y-\omega_{1}s)}{\lambda+k_j^{2}\omega_{1}^{*}}\\    \notag
&-\frac{4k_{j}^{2}|\omega_{1}|^{2}}{|\lambda+k_j^{2}\omega_{1}|^2}                                                 )+O(\varepsilon^{2},\varepsilon^{*2},|\varepsilon|^{2}),~~(j=1,2).
\end{align}
Taking the limit $\varepsilon\rightarrow 0$, and letting $\omega_{1}=\mu+i\upsilon$, the rogue wave is written as
\begin{equation}\label{74}
q_j=\tau_{j}e^{i\psi_{j}}(1+4\frac{L}{M}),
\end{equation}
where
\begin{align*}
L&=\lambda(k_j\upsilon)^2(\mu^2+\upsilon^2)-
i\lambda k_j\upsilon^2((\lambda\mu+k_{j}^2(\mu^2+\upsilon^2))(y-\mu s)+\lambda\upsilon^2 s),\\
M&=(\lambda^2+2\lambda\mu k_j^2+k_j^4(\mu^2+\upsilon^2))(\mu^3+\upsilon^2(\mu-\lambda(y-s\mu)^2)-\lambda\upsilon^4s^2),\\
x&=\lambda(y+s)-\frac{4\upsilon^{2}((\mu^2+\upsilon^2)s-\mu y)}{\upsilon^2(\lambda(y-\mu s)^2-\mu)+\lambda \upsilon^4s^2},~t=-s(j=1,2).
\end{align*}
the real number $\mu, \upsilon$ can still be determined by $\omega_{1}$ from\eqref{72}.
\begin{align*}
&\lambda^4+(\lambda k_2)^2(2\lambda\mu+(\mu^2-\upsilon^2)k_2^2-|\tau_{2}|^2)+k_{1}^2(\lambda^2(2\lambda\mu+|\tau_1|^2)
+k_2^4(2\lambda\mu(\mu^2-3\upsilon^2)+(\mu^2-\upsilon^2)|\tau_1|^2)\\
&+2\lambda k_2^2(2\lambda(\mu^2-\upsilon^2)+\mu(|\tau_1|^2-|\tau_2|^2)))
+k_1^4(\lambda^2(\mu^2-\upsilon^2)+k_2^4(\mu^4-6\mu^2\upsilon^2+\upsilon^4)\\
&+k_2^2(2\lambda\mu(\mu^2-3\upsilon^2)+(\upsilon^2-\mu^2)|\tau_2|^2))=0,\\
&2\upsilon(\lambda^2k_2^2(\lambda+\mu k_2^2)+k_1^2(\lambda^3+k_2^4
(3\lambda\mu^2-\lambda\upsilon^2+\mu|\tau_{1}|^2)+\lambda k_2^2(4\lambda\mu+|\tau_{1}|^2-|\tau_{2}|^2))\\
&+k_1^4(\lambda^2\mu+2\mu k_2^4(\mu^2-\upsilon^2)+k_2^2(3\lambda\mu^2-\lambda\upsilon^2-\mu|\tau_{2}|^2)))=0.
\end{align*}

The analytical expression of the rogue wave is given with six parameters. Through the direct calculation,
we can find that there exist four extreme points. The amplitude of the central point is
\begin{equation*}
|q_{j}(0,0)|^{2}=|\tau_j|^2
\frac{(\lambda^2\mu+2\lambda k_{j}^{2}(\mu^2+2\upsilon^2)
+\mu k_{j}^{4}(\mu^2+\upsilon^2))^2}{\mu^2(\lambda^2+2\lambda\mu k_{j}^{2}+k_{j}^{4}(\mu^2+\upsilon^2))^2},~(j=1,2).
\end{equation*}
We ignore the analysis of the other extreme points as the formulas are tedious and complicated.
The bright rogue wave and dark rogue wave are described in Fig.15(a1)and Fig.15(b1) respectively
and the four-petal rogue wave is also received in Fig.15(a2)-(b2). It should be remark here that
the dark rogue wave has not been reported in the short pulse equation.\\
\begin{figure}[!h]
\centering
\begin{minipage}{0.3\linewidth}\centering
\includegraphics[width=1\textwidth]{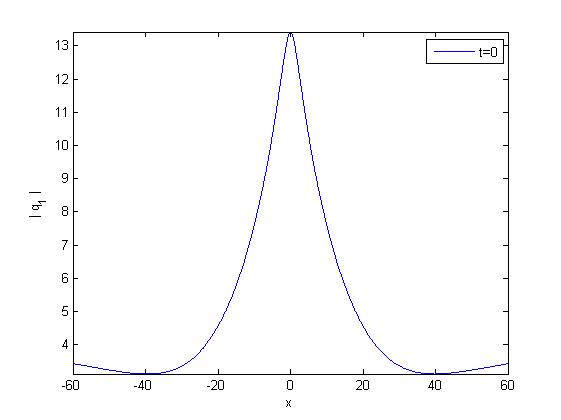}\caption*{(a1)}
\end{minipage}
\begin{minipage}{0.3\linewidth}\centering
\includegraphics[width=1\textwidth]{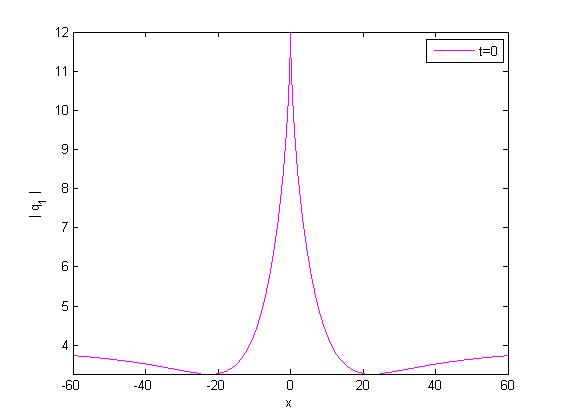}\caption*{(a2)}
\end{minipage}
\begin{minipage}{0.3\linewidth}\centering
\includegraphics[width=1\textwidth]{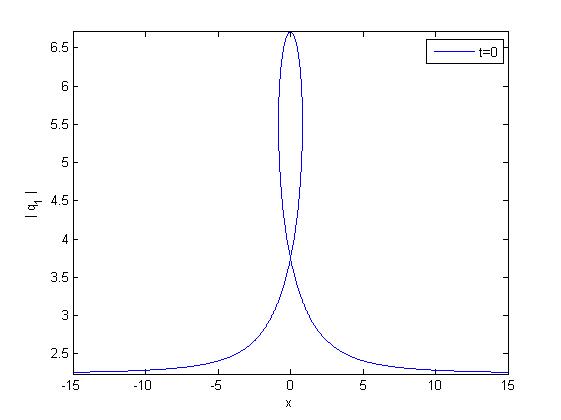}\caption*{(a3)}
\end{minipage}\\
\begin{minipage}{0.3\linewidth}\centering
\includegraphics[width=1\textwidth]{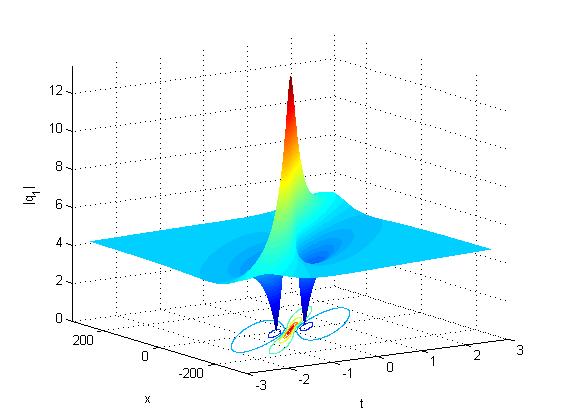}\caption*{(b1)}
\end{minipage}
\begin{minipage}{0.3\linewidth}\centering
\includegraphics[width=1\textwidth]{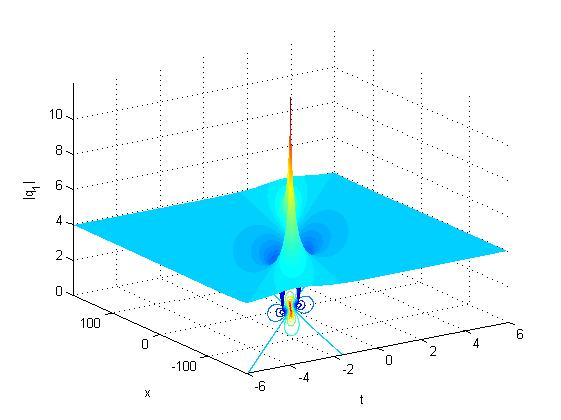}\caption*{(b2)}
\end{minipage}
\begin{minipage}{0.3\linewidth}\centering
\includegraphics[width=1\textwidth]{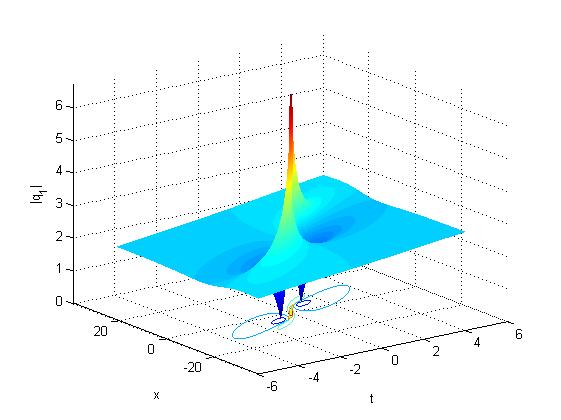}\caption*{(b3)}
\end{minipage}\\
\caption{\small{Rogue wave for Eq.\eqref{9} with 2D and 3D plots,
(a1)-(c1) smooth rogue wave soliton with parameter $\tau_{1}=2\sqrt{5}, \tau_{2}=\sqrt{7}, k=1$,
(a2)-(c2) cusponed rogue wave soliton with parameter $\tau_{1}=4, \tau_{2}=2, k=1$,
(a3)-(c3) loop rogue wave soliton with parameter $\tau_{1}=\sqrt{5}, \tau_{2}=\sqrt{2}, k=1$.}}
\label{fig16}
\end{figure}
To analyze the dynamics of the rogue wave solution for the Eq.\eqref{9}, we need
to solve the relation between $(x,t)$ and $(y,s)$. It is not
possible in general to give such a relation because of too many parameters involved.
However, we can obtain the relation for the reduced case.
By some tedious calculations, the single rogue wave solution can be constructed
from the formular \eqref{70a}-\eqref{71} by using the technique\cite{33}
\begin{align}  \notag
q_1&=\tau_{1}e^{i\theta}(1-4\frac{L}{M}),~~~~
q_2=\tau_{2}e^{i\theta}(1-4\frac{L}{M}), \\   \notag
M&=(|\tau_{1}|^{2}-|\tau_{2}|^{2})^{2}s^{2}+4k^{2}(1+k^{2}y^{2})+(|\tau_{1}|^{2}-|\tau_{2}|^{2})(4k^{2}s(s-y)+1),\\ \label{75}
L&=4k^{2}(1-iky)+(|\tau_{1}|^{2}-|\tau_{2}|^{2}),\\  \notag
x&=\frac{|\tau_{2}|^{2}-|\tau_{1}|^{2}}{2}(y+s)-4\frac{2k^{2}(2s-y)+
(|\tau_{1}|^{2}-|\tau_{2}|^{2})s}{(|\tau_{1}|^{2}-|\tau_{2}|^{2})s^{2}+
4k^{2}(1+k^{2}y^{2})/(|\tau_{1}|^{2}-|\tau_{2}|^{2})+4k^{2}s(s-y)+1},~t=-s.
\end{align}
which is nothing but a rogue wave solution of Eq.\eqref{9}. We emphasize
that the restriction $|\tau_{1}|>|\tau_{2}|$ in \eqref{75} provides the rogue wave
solution of the Eq.\eqref{9} which is localized in both space and time.
A typical evolution of the rogue wave
 is shown in Fig.16. The peak of the rogue wave \eqref{75} at point $(x,t)=(0,0)$ is
\begin{equation}\label{76}
|q_{1}|=3|\tau_{1}|,~|q_{2}|=3|\tau_{2}|,
\end{equation}
where the hole of the rogue wave \eqref{75} at point $(x,t)=(\pm\sqrt{\frac{3}{|\tau_{1}|^{2}-|\tau_{2}|^{2}}}, \pm\frac{3\sqrt{3}}{2}\sqrt{|\tau_{1}|^{2}-|\tau_{2}|^{2}})$ is zero.
It is interesting to analyze the role of the parameter $k$ in
the rogue wave solution as well. When we increase the value of $k$, the width of the profile increases. Further,
if $|\tau_{1}|^{2}-|\tau_{2}|^{2}>12k^{2}$ , one can obtain the smooth rogue wave solution (Figs.16(a1));
If $|\tau_{1}|^{2}-|\tau_{2}|^{2}=12k^{2}$, one obtains the cuspon rogue wave (Figs.16(a2));
If $|\tau_{1}|^{2}-|\tau_{2}|^{2}<12k^{2}$, one can obtain the loop rogue wave (Figs.16(a3)).
\section{Conclusions}
In this paper, we focus our attention on the coupled focusing-defocusing complex short pulse equation\eqref{9}.
By using the Hirota bilinear technique\cite{34}, the bright-bright, bright-dark,
dark-dark soliton solutions, breather and rogue wave solution for the Eq.\eqref{9} are constructed.
It has been shown that all these solitons have three shapes: smooth, cuspon, and
loop soliton.
The dynamics and asymptotic behavior of the solitons are analyzed.
The interaction of the two bright-bright solitons undergo energy exchanged collision
depending on the choices of the parameters.
However, the  interactions of bright-dark and dark-dark solitons are elastic.
We have shown that the
coupled focusing-defocusing complex short pulse equation admits the dark-dark soliton.
Starting from the dark-dark soliton solution,
we find two kinds of solutions: breather and rogue wave.
The breather can be expressed as trigonometric and hyperbolic functions.
The first order rational rogue wave solution
is received by resorting to the Taylor series
expansion coefficients. To the best of our knowledge, the dark rogue wave is found for the first time and deserves a further investigation.
We have seen that there exist some difference of the property of solution between coupled nonlinear Schr\"odinger equations\cite{25,26,27,28,29,30,31,32,33} and the coupled focusing-defocusing complex short pulse equation \eqref{9}.
\vskip 16pt \noindent {\bf
Acknowledgements} \vskip 12pt
The work of ZNZ is supported by the National Natural Science
Foundation of China under grant 11671255 and
by the Ministry of Economy and Competitiveness of Spain under
contract MTM2016-80276-P (AEI/FEDER,EU). We are very grateful to Prof. B.F. Feng for his many useful and constructive discussions.\\

\small{

}

\begin{thebibliography}{aa}
\bibitem{1}A. Hasegawa, Y. Kodama, Solitons in Optical Communications, Oxford University Press, New York, 1995.
\bibitem{2}G.P. Agrawal, Nonlinear Fiber Optics, Academic, San Diego, 2001.
\bibitem{3}V. G. Makhankov, Soliton Phenomenology, Kluwer Academic, London, 1990.
\bibitem{4}M. J. Ablowitz, Nonlinear Dispersive Waves: Asymptotic Analysis
and Solitons, Cambridge University Press, Cambridge, England, 2011.
\bibitem{5}M. J. Ablowitz, G. Biondini, and A. Lev Ostrovsky, Optical solitons: Perspectives and applications,  Chaos 10, 471 (2000).
\bibitem{6}J. C. Bronski, L. D. Carr, B. Deconinck, and J. N. Kutz, Bose-Einstein condensates
in standing waves: the cubic nonlinear Schr\"odinger equation with a periodic potential, Phys. Rev. Lett. 86, 1402 (2001).
\bibitem{7}E. P. Bashkin and A. V. Vagov, Instability and stratification of a two-component Bose-Einstein condensate
in a trapped ultracold gas, Phys. Rew. B 56, 6207 (1997).
\bibitem{8}J.E. Rothenberg, Space-time focusing: breakdown of the slowly varying envelope approximation in the self-focusing of femtosecond pulses, Opt. Lett. 17, 1340 (1992).
\bibitem{9}T. Sch\"afer, C.E. Wayne, Propagation of ultra-short optical pulses in cubic nonlinear media. Phys. D 196, 90-105 (2004).
\bibitem{10} A. Sakovich, S. Sakovich, The short pulse equation is integrable. J. Phys. Soc. Jpn. 74, 239-241 (2005).
\bibitem{11}J. C. Brunelli, The bi-hamiltonian structure of the short pulse equation, Phys. Lett. A 353, 475-478 (2006).
\bibitem{12}Y. Matsuno, Multiloop soliton and multibreather solutions of the short pulse model equation, J. Phys. Soc. Jpn. 76, 084003 (2007).
\bibitem{13}Y. Matsuno, Periodic solutions of the short pulse model equation, J. Math. Phys. 49, 073508 (2008).
\bibitem{14} R. Beals, M. Rabelo, K. Tenenblat, B\"acklund transformations and inverse scattering solutions for some pseudospherical surface equations, Stud. Appl. Math. 81, 125-151 (1989).
\bibitem{15}E. J. Parkes, Some periodic and solitary travelling-wave solutions of the short-pulse equation, Chaos, Solitons Fractals, 38, 154-159 ( 2008).
\bibitem{16}V. K. Kuetche, T. B. Bouetou, T. C. Kofane, On two-loop soliton solution of the sch\"afer-wayne short-pulse equation using hirota¡¯s method and hodnett¨Cmoloney approach. J. Phys. Soc. Jpn. 76, 024004 (2007).
\bibitem{17}B. F. Feng, K. Maruno, Y. Ohta, Integrable discretization of the short pulse equation, J. Phys. A 43, 085203 (2010).
\bibitem{18}B. F. Feng, J. Inoguchi, K. Kajiwara, K. Maruno, Y. Ohta, Discrete integrable systems and hodograph
transformations arising from motions of discrete plane curves, J. Phys. A 44, 395201 (2011).
\bibitem{19}S. F. Shen, B. F. Feng, Y. Ohta, From the real and complex coupled dispersionless equations to the real and complex short pulse equations,
Stud. Appl. Math. 136, 64 (2016).
\bibitem{20}B. F. Feng, Complex short pulse and coupled complex short pulse equations, Phys. D 297, 62-75 (2015).
\bibitem{21}L. Lin, B. F. Feng, Z. Zhu, Multi-soliton, multi-breather and higher order rogue wave solutions to the complex short pulse equation, Phys. D 327, 13-29 (2016).
\bibitem{22}B. F. Feng, L. Lin, Z. Zhu, Defocusing complex short-pulse equation and its multi-dark-soliton solution, Phys. Rew. E 93, 052227 (2016).
\bibitem{23}B. Guo, Y. F. Wang, Bright-dark vector soliton solutions for the coupled complex short pulse equations in nonlinear optics, Wave Motion 67, 47-54 (2016).
\bibitem{24}B. F. Feng, Private Communication (2016).
%\bibitem{25}S. Amiranashvili, A.G. Vladimirov, U. Bandelow, Few-optical-cycle solitons and
%pulse self-compression in a Kerr medium, Phys. Rev. A 77, 063821 (2008).
\bibitem{25}M. Vijayajayanthi, T. Kanna, and M. Lakshmanan, Bright-dark solitons and their collisions in mixed N-coupled nonlinear Schr\"odinger equations. Phy. Rev. A 77, 013820 (2008).
\bibitem{26}R. Radhakrishnan, M. Lakshmanan, J. Hietarinta, Inelastic collision and switching of coupled bright solitons in optical fibers, Phys. Rev. E 56, 2213 (1997).
\bibitem{27}T. Kanna, M. Lakshmanan, Exact soliton solutions, shape changing collisions, and partially coherent solitons in coupled nonlinear Schr\"odinger equations, Phys. Rev. Lett. 86, 5043 (2001).
\bibitem{28}T. Kanna, M. Lakshmanan, Exact soliton solutions of coupled nonlinear Schr\"odinger equations: shape-changing collisions, logic gates, and partially coherent solitons, Phys. Rev. E 67, 046617 (2003).
\bibitem{29}D. S. Wang, D. J. Zhang, and J. Yang, Integrable properties of the general coupled nonlinear Schr\"odinger equations
J. Math. Phys. 51, 023510 (2010).
\bibitem{30}F. Baronio, A. Degasperis, M. Conforti, and S. Wabnitz, Solutions of the vector nonlinear Schr\"odinger equations:
evidence for deterministic rogue waves, Phys. Rev. Lett. 109, 044102 (2012).

\bibitem{31}L. M. Ling, L. C. Zhao, and B. L. Guo, Darboux transformation and multi-dark soliton for N-component coupled nonlinear Schr\"odinger equations, Nonlinearity, 28, 3243-3261 (2015).
\bibitem{32}A. Mahalingam and K. Porsezian, Propagation of dark solitons in a system of coupled higher-order nonlinear
Schr\"odinger equations, J. Phys. A: Math. Gen. 35, 3099 (2002).
\bibitem{33}T. Kanna and M. Lakshmanan, Exact soliton solutions of coupled nonlinear Schr\"odinger equations: Shape-changing collisions,
logic gates, and partially coherent solitons, Phys. Rew. E 67, 046617 (2003).
\bibitem{34}R. Hirota, The Direct Method in Soliton Theory, Cambridge Univ. Press, Cambridge, 2004.
\bibitem{35}N. Vishnu Priya, M. Senthilvelan, and M. Lakshmanan, Dark solitons, breathers, and rogue wave solutions of the coupled generalized nonlinear Schr\"odinger equations, Phys. Rew. E 89, 062901 (2014).
\bibitem{36}Nail Akhmediev, Adrian Ankiewicz, and J. M. Soto-Crespo, Rogue waves and rational solutions of the nonlinear Schr\"odinger equation, Phys. Rev. E 80, 026601 (2009).
\bibitem{37}P. Dubard, P. Gaillard, C. Kleina, and V.B. Matveev, On multi-rogue wave solutions of the NLS equation
and positon solutions of the KdV equation, Eur. Phys. J. Special Topics 185, 247-258 (2010).
\bibitem{38}B. Guo, L. Ling, Rogue wave, breathers and bright-dark-rogue solutions for the coupled Schr\"odinger equations, Chin. Phys. Lett. 28, 110202 (2011).
\bibitem{39}B. Guo, L. Ling, Q.P. Liu, Nonlinear Schr\"odinger equation: generalized Darboux transformation and rogue wave solutions, Phys. Rev. E 85, 026607 (2012).
\bibitem{40}Y. Ohta and J. Yang, General high-order rogue waves and their dynamics in the nonlinear Schr\"odinger equation, Proc. R. Soc. A 468, 1716 (2012).

\end{thebibliography}
\end{document}